\documentclass[11pt,a4paper]{article}

\usepackage[T1]{fontenc}
\usepackage{lmodern}
\usepackage{microtype}
\usepackage{amsmath,amssymb,amsfonts,amsthm}
\usepackage{mathtools}
\usepackage{bm}
\usepackage{textcomp}
\usepackage{graphicx}
\usepackage{algorithm}
\usepackage{algpseudocode}
\usepackage{booktabs}
\usepackage{array}
\usepackage{xcolor}
\usepackage{etoolbox}
\usepackage{cite}
\usepackage[font=small,labelfont=bf,labelsep=period,figurename=Fig.,tablename=Table]{caption}
\usepackage[margin=2.5cm,headsep=0.8cm]{geometry}

\usepackage{titlesec}
\titleformat{\section}{\large\bfseries}{\thesection}{0.8em}{}
\titleformat{\subsection}{\normalsize\bfseries}{\thesubsection}{0.8em}{}
\titleformat{\subsubsection}[runin]{\normalsize\bfseries}{\thesubsubsection}{0.5em}{}[.]

\usepackage{fancyhdr}
\fancypagestyle{first}{\fancyhf{}\fancyfoot[C]{\small\thepage}}

\usepackage[colorlinks,linkcolor=black,citecolor=blue!50!black,urlcolor=blue!50!black]{hyperref}

\newtheorem{theorem}{Theorem}
\newtheorem{proposition}{Proposition}
\newtheorem{lemma}{Lemma}
\newtheorem{corollary}{Corollary}
\theoremstyle{definition}
\newtheorem{assumption}{Assumption}
\newtheorem{definition}{Definition}
\newtheorem{remark}{Remark}

\usepackage[most]{tcolorbox}
\tcbset{resstyle/.style={enhanced,breakable,colback=white,boxrule=0pt,frame hidden,arc=0pt,
  borderline west={1.6pt}{0pt}{black!55},left=7pt,right=6pt,top=3pt,bottom=3pt,before skip=9pt,after skip=9pt}}
\tcbset{defstyle/.style={enhanced,breakable,colback=white,boxrule=0pt,frame hidden,arc=0pt,
  borderline west={0.8pt}{0pt}{black!45},left=7pt,right=0pt,top=1pt,bottom=1pt,before skip=8pt,after skip=8pt}}
\tcolorboxenvironment{theorem}{resstyle}
\tcolorboxenvironment{lemma}{resstyle}
\tcolorboxenvironment{proposition}{resstyle}
\tcolorboxenvironment{corollary}{resstyle}
\tcolorboxenvironment{definition}{defstyle}
\tcolorboxenvironment{assumption}{defstyle}
\titleformat{\section}{\large\bfseries}{\thesection}{0.8em}{}[\vspace{-0.6ex}\textcolor{black!35}{\rule{\textwidth}{0.4pt}}]

\newcommand{\R}{\mathbb{R}}
\newcommand{\norm}[1]{\left\lVert#1\right\rVert}
\newcommand{\inner}[2]{\langle #1,\,#2\rangle}
\newcommand{\smin}{\sigma_{\min}}
\newcommand{\smax}{\sigma_{\max}}
\newcommand{\lmin}{\lambda_{\min}}
\newcommand{\lmax}{\lambda_{\max}}
\newcommand{\trans}{^{\!\top}}
\newcommand{\ddx}{\partial_x}
\newcommand{\ddth}{\partial_\theta}
\newcommand{\bzero}{\bm 0}
\newcommand{\bI}{\bm I}
\newcommand{\nx}{n_x}
\newcommand{\ny}{n_y}
\newcommand{\nth}{n_\theta}
\newcommand{\blkdiag}{\operatorname{blkdiag}}
\newcommand{\fk}[1]{f^{[#1]}}
\newcommand{\Mset}{\mathcal{M}}
\newcommand{\Wset}{\mathcal{W}}
\newcommand{\Jr}{J}
\newcommand{\Ups}{\Upsilon}
\newcommand{\eqdef}{\vcentcolon=}
\DeclareMathOperator*{\argmin}{arg\,min}

\newcommand{\appendices}{\clearpage\appendix
  \fancyhead[R]{\small\itshape Appendix}
  \begin{center}{\Large\bfseries Appendix\par}\end{center}\vspace{0.5em}}
\let\IEEEincludegraphics\includegraphics
\renewcommand{\includegraphics}[2][]{\begingroup\setlength{\columnwidth}{0.7\textwidth}%
  \IEEEincludegraphics[#1]{#2}\endgroup}

\begin{document}

\thispagestyle{first}
\begin{center}
{\small\itshape Preprint. Submitted to \emph{IEEE Transactions on Automatic Control}. November 2026.\par}
\vspace{1.8em}
{\LARGE\bfseries Overparameterized Simulation-Error Minimization for Nonlinear System Identification: Horizon-Uniform Landscape and Identifiability\par}
\vspace{1.5em}
{\large Cesare Donati\textsuperscript{a,*}, Constantino Lagoa\textsuperscript{b}\par}
\vspace{0.9em}
{\small\textsuperscript{a}\,Istituto di Elettronica e di Ingegneria dell'Informazione e delle Telecomunicazioni, Consiglio Nazionale delle Ricerche (CNR-IEIIT), c/o Politecnico di Torino, Corso Duca degli Abruzzi 24, 10129 Torino, Italy\par\smallskip
\textsuperscript{b}\,EECS, The Pennsylvania State University, University Park, PA, USA\par}
\end{center}
\vspace{1.2em}
{\small\leftskip=1cm\rightskip=1cm\noindent\textbf{Abstract.}
Simulation-error minimization is a standard criterion in nonlinear system
identification, but its landscape is nonconvex and its conditioning can
degrade geometrically with the data horizon $N$. {We propose to lift the trajectory with a free, softly penalized per-step correction, which yields a fit loss that gradient descent provably minimizes:} under a rank condition on the output
map and a bound on the one-step Jacobian, it satisfies a
Polyak--{\L}ojasiewicz inequality with an explicit constant independent of
$N$. A zero-loss model leaves the physical parameter undetermined.
Identifiability is restored by a minimal-correction selection principle: under
structural identifiability, the global minimizer of the penalized loss returns
the true parameters for every positive penalty weight, and under noise and
unmodeled dynamics the parameter error grows at most linearly with the
mismatch. {The resulting estimator, gradient descent on the fit loss followed by a
homotopy in the penalty weight, reconciles fit and selection, with a local convergence guarantee. In experiments, it recovers the parameters on long records of expanding and chaotic systems on which single shooting fails.}

\par}
\vspace{1.5em}
\begingroup\renewcommand{\thefootnote}{}\footnotetext{\hspace{-1.8em}This work has been submitted to the IEEE for possible publication. Copyright may be transferred without notice, after which this version may no longer be accessible.\par\noindent\hspace{-1.8em}\textsuperscript{*}\,Corresponding author: \href{mailto:cesaredonati@cnr.it}{cesaredonati@cnr.it}. C.~Lagoa: \href{mailto:cml18@psu.edu}{cml18@psu.edu}.}\endgroup

\section{Introduction}
\label{sec:introduction}

Decisions in engineering and science increasingly rest on dynamical models
\emph{learned} from measured data, from predictive control and condition
monitoring to the digital twins that mirror physical assets. Constructing
such a model, i.e., the problem of \emph{system identification}, is therefore
foundational. For linear time-invariant systems the theory is
mature~\cite{ljung1999system, van1994n4sid}; for \emph{nonlinear} systems,
which describe most real processes, identification remains open in
fundamental ways~\cite{ljung2010perspectives, sjoberg1995nonlinear}. A model
that must predict, simulate, or be controlled over a horizon has to
reproduce the measured trajectory when run forward on its own from a single
initial condition. This requirement is formalized by the \emph{simulation
error}, the discrepancy between the record and the free-run output of the
model, which is the criterion adopted, e.g.,
in~\cite{ribeiro2020smoothness, beintema2021encoder}.
The classical one-step-ahead prediction error~\cite{ljung1999system} asks
less: a model can make it small and still drift away from the data once it is
iterated on its own outputs. Minimizing simulation error is, however,
demanding~\cite{farina2011simulation}. Each simulated output depends on the
parameters through a composition of the model as long as the elapsed time, so
that, where the model is not contractive, the bounds on the Lipschitz and
smoothness constants of the cost grow exponentially with the length of the
record, or horizon, and the cost can exhibit many local
minima~\cite{ribeiro2020smoothness}.

Lifting the trajectory improves this landscape in practice, but none of the
lifted schemes guarantees that gradient-based estimation reaches a global
minimizer. In multiple shooting, developed for optimal
control~\cite{bock1984multiple} and brought to identification to make the
simulation length a design parameter~\cite{ribeiro2020smoothness}, the record
is split into short segments, the states at which the segments start become
additional decision variables, and continuity between consecutive segments is
imposed as a constraint. Related schemes replace the full simulation by
multi-step predictions~\cite{farina2011simulation} or estimate the initial
state of each short section with an encoder network~\cite{beintema2021encoder},
and segmentation makes it possible to fit neural differential equations to
oscillatory records on which single shooting fails~\cite{turan2022multiple}
and to train them stably on long trajectories~\cite{iakovlev2023latent}.
Continuity is enforced exactly~\cite{bock1984multiple, ribeiro2020smoothness}
or relaxed through a penalty or an augmented
Lagrangian~\cite{turan2022multiple} or a continuity-inducing
prior~\cite{iakovlev2023latent}, and the benefit rests on smoothness
arguments~\cite{ribeiro2020smoothness} and on experiments.

Global guarantees without convexity are available, instead, for
overparameterized models. The Polyak--\L ojasiewicz (PL) inequality, which
bounds the suboptimality of a cost by its squared gradient norm, makes every
stationary point a global minimizer and yields linear convergence of gradient
methods~\cite{polyak1963gradient, lojasiewicz1963, karimi2016linear}. When a
model has more parameters than data to fit, gradient descent converges
globally provided the tangent kernel of the model stays positive
definite~\cite{jacot2018neural, oymak2020toward}, a condition that yields a PL
inequality~\cite{liu2022loss}. For recurrent networks, which unroll a model
over the horizon as simulation error does, a PL inequality near a random
initialization gives linear convergence of gradient descent to zero training
loss once the width grows polynomially with the sequence
length~\cite{allenzhu2019convergence}. The global minimizers then generically form a
manifold of positive dimension~\cite{cooper2021global}, and which of them is
returned becomes a matter of selection, as in adaptive control, where the
adaptation law selects, among the many parameters that achieve perfect
tracking or prediction, one with prescribed properties~\cite{boffi2021implicit}.
In control, gradient domination of PL type underlies the global convergence of
policy optimization for the linear quadratic regulator~\cite{fazel2018global},
also under weaker forms of the inequality~\cite{deoliveira2025remarks}, and
gradient descent provably learns linear dynamical
systems~\cite{hardt2018gradient}. Simulation-error identification, however, is
not overparameterized: over a long record, the parameters and the initial
state are far fewer than the measured outputs they must reproduce.

{
This paper supplies the missing overparameterization by design. For a model
$x_{k+1}=f(x_k,\theta)$, $y_k=g(x_k)$ and a record $y_1,\dots,y_N$, we propose to
estimate $\theta$ and $x_0$ by solving
\begin{equation*}
\begin{aligned}
  \min_{\theta,\,x_0,\,z,\,d}\ \sum_{k=1}^{N}\Big(&\norm{z_k-f(z_{k-1},\theta)-d_k}^2\\
  &+\norm{y_k-g(z_k)}^2+\gamma\norm{d_k}^2\Big),
\end{aligned}
\end{equation*}
in which the trajectory is lifted to free states $z_k$, with $z_0=x_0$, and each
step receives a free correction $d_k$, penalized with weight $\gamma\ge0$. The
estimate is computed in two phases: gradient descent at $\gamma=0$ fits the data,
and a continuation that decreases $\gamma$ to a target value then selects one of
the exact fits. The equations $z_k=f(z_{k-1},\theta)+d_k$ are as many as the components of the lifted states, and the equations $g(z_k)=y_k$ are at most as many as the components of the corrections whenever the state has at least as many components as the output. For every horizon, the lifted problem then has more unknowns than equations, with $\theta$ and $x_0$ in excess.
}
Under a rank condition on the output map and a bound on the
one-step state Jacobian, the resulting fit loss satisfies a PL inequality with
a constant independent of the horizon on the region where the two conditions
hold, so gradient descent converges linearly to a model that reproduces the
data exactly whenever its iterates remain in that region. An exact fit is thus
easy by construction, and the substance of the problem moves to selection.

The price of this freedom is identifiability: an exact fit no longer
determines the physical parameter. The same tension arises when a physical
model is augmented with a flexible component, where a penalty on the
component, its norm in a reproducing kernel Hilbert
space~\cite{donati2026kernel} or a sparsity-promoting penalty on the weights of
a basis-function expansion~\cite{donati2025combining}, arbitrates between the
two parts. We restore identifiability by selecting, among the exact fits, the
one with minimal correction. Under structural identifiability, this selection
returns the true parameters, and the global minimizer of the penalized loss
realizes it at every positive penalty weight; under noise and unmodeled
dynamics, the parameter error grows at most linearly with the mismatch.
Unlike the fit, the selection is not global for gradient methods, since the
penalized loss can have spurious local minimizers; a homotopy in the penalty
weight reconciles fit and selection, with a local convergence guarantee. To our
knowledge, no prior work establishes a PL inequality with a horizon-uniform
constant for simulation-error identification, nor shows that minimizing the
correction restores identifiability of the physical parameter on the lifted
landscape.

\subsubsection*{Outline}
Section~\ref{sec:simerror} formalizes simulation-error minimization, shows
why its landscape is ill-conditioned and nonconvex, and states the two
questions answered in the paper. Section~\ref{sec:formulation} introduces a
lifted reformulation with a free, penalized correction, whose fit loss
satisfies a PL inequality with a horizon-independent constant and whose
conditioning degrades at most linearly in the horizon
(Section~\ref{sec:geometry}). Section~\ref{sec:curse} shows that the
reformulation alone leaves the parameters undetermined, and
Section~\ref{sec:selection} restores identifiability through a
minimal-correction selection principle, with exact recovery, a continuation
scheme that starts selection from the output of the fit, and
robustness linear in the unmodeled effects. Sections~\ref{sec:numerics} and~\ref{sec:conclusions} report numerical evidence and conclusions.

\subsubsection*{Notation}
$\norm{\cdot}$ is the Euclidean norm and the induced spectral norm. For a
matrix $M\in\R^{n\times m}$, $\smin(M)$ and $\smax(M)=\norm{M}$ are its
smallest and largest singular values, the former being the $\min\{n,m\}$-th,
and $\kappa(M)\eqdef\smax(M)/\smin(M)$ is its condition number; $\lmin(\cdot)$
and $\lmax(\cdot)$ are the extreme eigenvalues of a symmetric matrix,  and $\succeq$ is the Loewner order on symmetric matrices. $\bzero$
and $\bI$ are zero and identity matrices, with a subscript indicating the
size when needed. For a differentiable map $\psi$, $\partial\psi$ is its
Jacobian and $\partial_x\psi$ its partial Jacobian with respect to $x$. For a matrix $M$, $\ker M\eqdef\{v:Mv=0\}$ is its null space.
Vectors $v_1,\dots,v_N\in\R^n$ are stacked as $v=(v_1,\dots,v_N)\in\R^{Nn}$,
$[X_k]_{k=a}^b$ is the vertical stacking of the blocks $X_a,\dots,X_b$, and
$\blkdiag(M_1,\dots,M_k)$ is the block-diagonal matrix with diagonal blocks
$M_1,\dots,M_k$. For a set $\mathcal S$, $\operatorname{int}\mathcal S$ is its
interior. For a scalar $t$, $t\downarrow0$ denotes $t\to0$ with $t>0$.

\subsubsection*{List of symbols}
The analysis involves several groups of variables and constants, each
introduced where it is first needed; for reference, they are collected in
Table~\ref{tab:notation} in Appendix~\ref{app:notation}, with the place where each is defined.

\section{Simulation-error minimization}
\label{sec:simerror}
Before formulating the problem, we make precise the obstacles mentioned in
the Introduction, which determine what the formulation must achieve.
Suppose that the measurements $y_1,\dots,y_N$ are to be explained by a
parametric state-space model
\begin{subequations}\label{eq:system}
\begin{align}
  x_{k+1} &= f(x_k,\theta), \label{eq:system-state}\\
  y_k     &= g(x_k),         \label{eq:system-output}
\end{align}
\end{subequations}
where $x_k\in\R^{\nx}$ is the unmeasured state, $y_k\in\R^{\ny}$ the measured
output, $\theta\in\R^{\nth}$ the unknown parameter, and
$f:\R^{\nx}\times\R^{\nth}\to\R^{\nx}$ and $g:\R^{\nx}\to\R^{\ny}$ the state
transition and observation maps. Let $\fk{k}(x_0,\theta)$ denote the state
obtained by iterating $f(\cdot,\theta)$ exactly $k$ times from $x_0$, and let
\begin{equation}
  \Ups(\theta,x_0)\eqdef\big(g(\fk{1}(x_0,\theta)),\dots,g(\fk{N}(x_0,\theta))\big)
  \label{eq:Upsilon}
\end{equation}
be the simulated output trajectory, with blocks $\Ups_k(\theta,x_0)\eqdef g(\fk{k}(x_0,\theta))$. With $y\eqdef(y_1,\dots,y_N)\in\R^{N\ny}$,
the \emph{simulation error} is
\begin{equation}
  L_{\mathrm{sim}}(\theta,x_0){\eqdef}\norm{y-\Ups(\theta,x_0)}^2
  {=}\sum_{k=1}^{N}\norm{y_k-g\big(\fk{k}(x_0,\theta)\big)}^2\!.
  \label{eq:simerror}
\end{equation}
Each simulated output depends on $(\theta,x_0)$ through the composition
$\fk{k}$, and so does its sensitivity. Along the simulated trajectory
$x_k=\fk{k}(x_0,\theta)$, let $A_k\eqdef\ddx f(x_{k-1},\theta)$,
$B_k\eqdef\ddth f(x_{k-1},\theta)$ and $C_k\eqdef\ddx g(x_k)$,
$k=1,\dots,N$, and let $\Phi_{k,j}\eqdef A_k\cdots A_{j+1}$ for $0\le j<k$,
with $\Phi_{k,k}\eqdef\bI_{\nx}$. By the chain rule, the Jacobian of the
simulated output is
\begin{equation}
  \partial\Ups=
  \begin{bmatrix}
    C_k\sum_{j=1}^{k}\Phi_{k,j}B_j, & C_k\,\Phi_{k,0}
  \end{bmatrix}_{k=1}^{N}
  \in\R^{N\ny\times(\nth+\nx)},
  \label{eq:Jsim}
\end{equation}
whose column blocks are the sensitivities to $\theta$ and to $x_0$. With
$\underline a\eqdef\min_k\smin(A_k)$, submultiplicativity of the spectral norm
and $\smin(AB)\ge\smin(A)\smin(B)$ give
\begin{equation}
  \underline a^{\,k-j}\le\smin(\Phi_{k,j})\le\norm{\Phi_{k,j}}\le\big(\max_i\norm{A_i}\big)^{k-j}.
  \label{eq:phibounds}
\end{equation}
Perturbations of $x_0$, and of $\theta$ through the early steps, are
therefore amplified exponentially along the horizon if $\underline a>1$ and
attenuated exponentially if $\max_k\norm{A_k}<1$.

Two obstacles follow. First, the curvature of $L_{\mathrm{sim}}$ grows
geometrically with the horizon when the dynamics expand. The last block row of
the $x_0$-columns of \eqref{eq:Jsim} is $C_N\Phi_{N,0}$, and
$\norm{C_N\Phi_{N,0}}\ge\norm{C_N}\smin(\Phi_{N,0})$, so
\eqref{eq:phibounds} gives
\begin{equation}
  \norm{\partial\Ups}^2\ \ge\ \norm{C_N}^2\,\underline a^{\,2N}.
  \label{eq:cond}
\end{equation}
Here $\norm{\partial\Ups}^2$ is the largest eigenvalue of the Gauss--Newton
matrix $\partial\Ups\trans\partial\Ups$, which equals half the Hessian of
$L_{\mathrm{sim}}$ at a zero-residual minimizer. Near such a minimizer,
gradient descent is therefore stable only for steps below
$\norm{\partial\Ups}^{-2}$, which decay at least geometrically in $N$ if
$\underline a>1$. This lower bound complements the upper bounds
of~\cite{ribeiro2020smoothness}, which grow exponentially with $N$ when the
one-step Lipschitz constant of the model exceeds one. The smallest eigenvalue
need not grow at the same rate, so the condition number can grow geometrically
as well. Second, $L_{\mathrm{sim}}$
is in general nonconvex and may have nonglobal local minimizers, to which
gradient descent initialized away from the truth can
converge~\cite{ribeiro2020smoothness}; contraction, $\max_k\norm{A_k}<1$, removes the first obstacle but not the second.

These obstacles rule out a global guarantee for gradient descent on
$L_{\mathrm{sim}}$ along the classical routes. Global convergence of gradient
descent is guaranteed in two settings, convexity and
overparameterization, and simulation-error minimization belongs to neither. It is in general nonconvex, as noted above.
Nor is it overparameterized: in overparameterized learning, the PL inequality
follows from the Jacobian of the residual having full row
rank~\cite{liu2022loss}, which requires at least as many decision variables
as scalar equations to be fitted, whereas the residual $y-\Ups(\theta,x_0)$
of \eqref{eq:simerror} has Jacobian $-\partial\Ups$, with more rows than
columns, hence not of full row rank, as soon as $N\ny>\nth+\nx$. This leads to the two questions addressed
in the rest of the paper.
\begin{itemize}
  \item[\footnotesize (Q1)] Can the problem be \emph{reformulated} so that gradient descent
  provably converges to a global optimum, with a guarantee uniform in the horizon $N$?
  \item[\footnotesize (Q2)] If so, does solving the reformulated problem still recover the
  \emph{true physical parameters}?
\end{itemize}

\section{Problem formulation}
\label{sec:formulation}

Given the record $y\in\R^{N\ny}$, the task is to estimate the parameter
$\theta$ and the initial condition $x_0\in\R^{\nx}$ of \eqref{eq:system}. The
simulation-error estimator \eqref{eq:simerror} addresses this task directly,
but inherits the ill-conditioning and the nonglobal minimizers described in
Section~\ref{sec:simerror}. We now reformulate it, keeping the model
$f(\cdot,\theta)$ and its physical parameter $\theta$, in a form suited to
(Q1) and (Q2), under the following standard regularity condition.

\begin{assumption}[Smoothness]\label{ass:smooth}
The maps $f$ and $g$ are continuously differentiable, and their Jacobians are
Lipschitz continuous on every compact subset of their domains.
\end{assumption}
 
\subsection{The lifted overparameterized model}
\label{subsec:lifted}

To make the problem overparameterized while retaining $f(\cdot,\theta)$ and
$\theta$, the decision space is enlarged along two independent directions. First,
following the multiple-shooting philosophy~\cite{bock1984multiple,
ribeiro2020smoothness}, the simulated trajectory is
promoted to a free \emph{lifted state} $z=(z_1,\dots,z_N)\in\R^{N\nx}$, with
the convention $z_0\eqdef x_0$, and the dynamics enter the cost as residuals
rather than as constraints. Second, the dynamics are augmented by free per-step
\emph{corrections} $d=(d_1,\dots,d_N)\in\R^{N\nx}$, which the cost penalizes
instead of forcing them to zero. All
decision variables are collected in
\begin{equation}
  w \eqdef (\theta,\,x_0,\,z,\,d)\in\R^{p},
  \qquad p\eqdef\nth+\nx+2N\nx .
  \label{eq:param-vector}
\end{equation}

The dynamics and the output equation thus enter the cost through
the \emph{consistency} and \emph{output residuals}
\begin{align}
  e_k(w) &\eqdef z_k - f(z_{k-1},\theta) - d_k, & k&=1,\dots,N, \label{eq:res-e}\\
  s_k(w) &\eqdef y_k - g(z_k),                  & k&=1,\dots,N. \label{eq:res-s}
\end{align}
Stacking $e\eqdef(e_1,\dots,e_N)\in\R^{N\nx}$ and
$s\eqdef(s_1,\dots,s_N)\in\R^{N\ny}$ into $r(w)\eqdef(e(w),s(w))\in\R^{m}$,
$m\eqdef N(\nx+\ny)$, we define the \emph{fit loss} and the
\emph{correction-penalized loss}
\begin{subequations}
\begin{align}
  L_0(w)&\eqdef\norm{r(w)}^2,\label{eq:L0}\\
  L_\gamma(w)&\eqdef L_0(w)+\gamma\,\norm{d}^2,\label{eq:Lgam}
\end{align}
  \label{eq:losses}%
\end{subequations}
where the weight $\gamma\ge0$ penalizes \emph{only} the correction, while
$\theta$, $x_0$ and $z$ remain unpenalized. The two terms play asymmetric
roles: $L_0$ is the landscape that gradient descent must traverse and the
object of the optimization analysis, whereas $\gamma\norm{d}^2$ is a
\emph{selection} term that determines which model is returned when several
fit the data equally well.

Recall from Section~\ref{sec:simerror} that the row-rank route to a PL
inequality requires at least as many decision variables as
scalar equations to be fitted, here $p\ge m$, and that simulation-error
minimization violates this as soon as $N\ny>\nth+\nx$. The lifted problem
satisfies it whenever $\nx\ge\ny$: the corrections supply $N\nx$ decision variables matched by
\emph{no} residual equation, so that
\begin{equation}
  p-m \;=\; \nth + \nx + N(\nx-\ny) ,
  \label{eq:margin}
\end{equation}
which grows linearly in $N$ when $\nx>\ny$. Without the corrections, lifting is dimensionally
neutral: each lifted state brings $\nx$ variables and $\nx$ consistency
equations, and the margin is $\nth+\nx-N\ny$, exactly that of
\eqref{eq:simerror}. The penalized correction is thus the ingredient that
makes the reformulation overparameterized; the next section shows that this
surplus of variables yields a PL inequality with a horizon-independent
constant.

\begin{remark}[Role of the lifted state]\label{rem:keepz}
Given $(\theta,x_0,d)$, the recursion $z_k=f(z_{k-1},\theta)+d_k$ determines
$z$ uniquely. Eliminating $z$ in this way leaves the correction free, and
hence the problem overparameterized, but reinstates the composition $\fk{k}$
of \eqref{eq:simerror}: the output sensitivities then contain products of
state Jacobians as in \eqref{eq:Jsim}, and the curvature bound
\eqref{eq:cond} applies again.
Keeping $z$ free confines the dynamics to one-step Jacobians.
\end{remark}

\section{Optimization geometry}
\label{sec:geometry}
This section shows that, under a rank condition on the output map and a
bound on the one-step state Jacobian, the fit loss $L_0$ satisfies a PL
inequality with an explicit, horizon-independent constant on the whole region
where the two conditions hold.

\subsection{A PL lemma for least squares}
\label{subsec:pl-lemma}

A continuously differentiable function $L:\R^p\to\R$ with
$L^\star\eqdef\inf L>-\infty$ satisfies the PL inequality
\cite{polyak1963gradient, lojasiewicz1963} with constant $\mu>0$ on a set
$\Wset\subseteq\R^p$ if
$$\tfrac12\norm{\nabla L(w)}^2\ge\mu\,(L(w)-L^\star) \quad \forall w\in\Wset;$$
every stationary point in $\Wset$ is then a global minimizer, and gradient descent with step equal to the inverse Lipschitz modulus of $\nabla L$ converges linearly~\cite{karimi2016linear}. For least squares, the PL constant is controlled by
the \emph{tangent kernel}. {The definition and the lemma below concern a
generic residual map; the following subsections apply them to the lifted
residual $r=(e,s)$ of \eqref{eq:losses}, for which $L=L_0$.}
 
\begin{definition}[Residual Jacobian and tangent kernel]\label{def:kernel}
For a continuously differentiable residual map $r:\R^p\to\R^m$, let
$\Jr(w)\eqdef\partial r(w)\in\R^{m\times p}$. The \emph{tangent kernel} is the
positive semidefinite matrix $K(w)\eqdef \Jr(w)\,\Jr(w)\trans\in\R^{m\times m}$,
and $r$ is a \emph{submersion} on $\Wset$ if $\Jr(w)$ has full row rank for
every $w\in\Wset$.
\end{definition}
 
\begin{lemma}[Submersion implies PL]\label{lem:pl}
Let $r:\R^p\to\R^m$ be continuously differentiable and let $L=\norm{r}^2$. If there exists
$\mu>0$ such that $\lmin\!\big(K(w)\big)\ge \mu/2$ for all $w\in\Wset$, then $L$ satisfies
the PL inequality
\begin{equation}
  \tfrac12\,\norm{\nabla L(w)}^2 \;\ge\; \mu\,L(w),
  \qquad \forall\,w\in\Wset .
  \label{eq:pl}
\end{equation}
Moreover, $L\ge 0$, so \eqref{eq:pl} is the PL inequality with reference value
$L^\star=0$; in particular every stationary point in $\Wset$ satisfies $r=0$ and
is therefore a global minimizer of $L$ on~$\R^p$.
\end{lemma}
 
\begin{proof}
The gradient of $L=\norm r^2=r\trans r$ is $\nabla L=2\,\Jr\trans r$. Hence
\begin{equation}
\begin{aligned}
  \norm{\nabla L}^2 &= 4\,r\trans \Jr\Jr\trans r = 4\,r\trans K\,r
  \\&\ge\; 4\,\lmin(K)\,\norm{r}^2
  \;\ge\; 2\mu\,\norm{r}^2 = 2\mu\,L ,
  \end{aligned}
  \label{eq:pl-chain}
\end{equation}
where the first inequality is the Rayleigh bound for the symmetric matrix $K$ and the
second uses $\lmin(K)\ge\mu/2$. Dividing by $2$ gives \eqref{eq:pl}. If
$\nabla L(w)=0$ at some $w\in\Wset$, then \eqref{eq:pl} forces $L(w)=0$, i.e.\ $r(w)=0$.
\end{proof}
For $p\ge m$, $\lmin(K)=\smin(\Jr)^2$, so the hypothesis of
Lemma~\ref{lem:pl} states that $r$ is a submersion on $\Wset$ with
$\smin(\Jr)$ bounded away from zero uniformly. For the lifted residual, { with
$p=\nth+\nx+2N\nx$ decision variables and $m=N(\nx+\ny)$ residuals,}
$p\ge m$ holds whenever $\nx\ge\ny$, by \eqref{eq:margin}. The optimization
argument thus reduces to a uniform lower bound on $\smin(\Jr)$; to obtain it,
we first examine the structure of the lifted Jacobian.
 
\subsection{The residual Jacobian and its structure}
\label{subsec:jacobian}

At $w=(\theta,x_0,z,d)$, define the one-step sensitivities along the lifted
states,
\begin{equation}
  A_k\eqdef \ddx f(z_{k-1},\theta),\ \
  B_k\eqdef \ddth f(z_{k-1},\theta),\ \
  C_k\eqdef \ddx g(z_k),
  \label{eq:sens}
\end{equation}
$k=1,\dots,N$, which reduce to those of Section~\ref{sec:simerror} when $z$ is
the simulated trajectory, and the functions
\begin{equation}
  a(w)\eqdef \max_{1\le k\le N}\norm{A_k},
  \qquad
  c(w)\eqdef \min_{1\le k\le N}\smin(C_k).
  \label{eq:scalars}
\end{equation}
Ordering the rows as $(e,s)$ and the columns as $(\theta,x_0,z,d)$,
differentiation of \eqref{eq:res-e}--\eqref{eq:res-s} gives
\begin{equation}
  \Jr=
  \begin{bmatrix}
    B_e & X_e & D_f & -\bI_{N\nx}\\[2pt]
    \bzero & \bzero & -G & \bzero
  \end{bmatrix},
  \label{eq:Jacobian}
\end{equation}
where $B_e\eqdef-[B_k]_{k=1}^N\in\R^{N\nx\times\nth}$,
$X_e\in\R^{N\nx\times\nx}$ has $-A_1$ as its first block and zeros elsewhere,
$G\eqdef\blkdiag(C_1,\dots,C_N)\in\R^{N\ny\times N\nx}$, and
  $D_f\eqdef\partial_z e\in\R^{N\nx\times N\nx}$ is block lower-bidiagonal,
with identity diagonal blocks and subdiagonal blocks $-A_2,\dots,-A_N$.
Three structural features of \eqref{eq:Jacobian} make a uniform lower bound
on $\smin(\Jr)$ possible. The correction block $\partial_d e=-\bI_{N\nx}$ is
perfectly conditioned, so the consistency rows of $\Jr$ are linearly
independent whatever the dynamics. The output residual depends on $w$ only
through $z$, and its Jacobian $G$ is block diagonal, so that $\smin(G)=c(w)$
involves no product over the horizon. Finally, in the $z$-columns the dynamics
enter only through $D_f$, whose spectral norm admits a one-step bound, $\norm{D_f}\le1+a(w)$ independently of $N$ (Appendix~\ref{app:pf-thm-pl}).

\subsection{Global PL and its consequences}
\label{subsec:pl-theorem}
With this bound, the submersion hypothesis of Lemma~\ref{lem:pl}
reduces, for the structured Jacobian \eqref{eq:Jacobian}, to an
instantaneous output-rank condition and a uniform bound on the one-step
state Jacobians, which we state as assumptions over a region of interest
$\Wset\subseteq\R^p$.
 
\begin{assumption}[Output submersion] \label{ass:submersion}
The output dimension satisfies $\ny\le\nx$, and there exists $\underline c>0$ such
that {the function $c(w)=\min_k\smin(C_k)$ of \eqref{eq:scalars}} satisfies
$c(w)\ge\underline c$ for all $w\in\Wset$. In particular, each $C_k$ has full row rank.
\end{assumption}

\begin{assumption}[Bounded one-step sensitivity] \label{ass:bounded}
There exists $\bar a<\infty$ such that {the function $a(w)=\max_k\norm{A_k}$ of
\eqref{eq:scalars}} satisfies $a(w)\le\bar a$ for all $w\in\Wset$.
\end{assumption}
Assumption~\ref{ass:submersion} is a non-degeneracy condition on the output
map $g$ alone, independent of the dynamics $f$. For an affine output $g(x)=Cx+g_0$, such as a measurement of some state
coordinates, $C_k=C$ for every $k$, so the assumption holds on all of $\R^p$
whenever $C$ has full row rank, with $\underline c=\smin(C)$. For a scalar
output, $C_k=\nabla g(z_k)\trans$ is a row vector and $\smin(C_k)=\norm{\nabla g(z_k)}$,
so the assumption requires the gradient of $g$ to stay away from zero, by at
least $\underline c$, at the lifted states $z_k$ of every $w\in\Wset$. More
generally, if the lifted states remain in a compact set on which $\ddx g$ has
full row rank, continuity of $x\mapsto\smin(\ddx g(x))$ provides
$\underline c>0$; the requirement is then that they stay away from the
critical set of $g$, where $\ddx g$ loses row rank, which is empty for affine
outputs. The condition becomes \emph{easier} to satisfy as $\ny$
decreases: the sensor-poor regime,
the most demanding for identification, is thus the benign one for the optimization
geometry. Assumption~\ref{ass:bounded} holds on any compact $\Wset$ by
Assumption~\ref{ass:smooth}; it bounds the one-step state sensitivity only,
with no product over the horizon, and it holds on all of $\R^p$ when $\ddx f$
is bounded on $\R^{\nx}\times\R^{\nth}$. Neither condition involves stability
or observability of the model. Under the two conditions, the tangent kernel is
uniformly positive definite on $\Wset$, with a floor that depends only on
$(\underline c,\bar a)$, as the following theorem shows.
\begin{theorem}[Global PL with a horizon-uniform constant]\label{thm:pl}
Under Assumptions~\ref{ass:smooth},~\ref{ass:submersion},
and~\ref{ass:bounded}, one has $\lmin\big(K(w)\big)\ge\mu/2$ for all $w\in\Wset$ and
\begin{equation}
  \tfrac12\,\norm{\nabla L_0(w)}^2 \;\ge\; \mu\,L_0(w)
  \qquad \forall\,w\in\Wset ,
  \label{eq:pl-L0}
\end{equation}
with the explicit constant
\begin{equation}
  \mu = \min\!\left\{\frac{2(\Omega+1)}{2\Omega+1},\
                      \frac{\underline c^{\,2}}{\Omega+1}\right\},
  \qquad
  \Omega\eqdef(1+\bar a)^2 ,
  \label{eq:pl-const}
\end{equation}
which depends only on $(\underline c,\bar a)$ and not on the horizon $N$.
\end{theorem}
\begin{proof}See Appendix~\ref{app:pf-thm-pl}.\end{proof}

The bound holds pointwise: if $\ny\le\nx$, then for every $w$ with $c(w)>0$, the proof in Appendix~\ref{app:pf-thm-pl} gives
$\lmin(K(w))\ge\mu(w)/2$, where $\mu(w)$ denotes \eqref{eq:pl-const}
evaluated at $\underline c=c(w)$ and $\bar a=a(w)$. Since
\eqref{eq:pl-const} decreases in $\bar a$ and increases in $\underline c$,
$\mu(w)\ge\mu$ on $\Wset$.

\begin{remark}[Reading the constant]\label{rem:reading}
The second branch of \eqref{eq:pl-const} is active if and only if
$\underline c^{\,2}(2\Omega+1)\le 2(\Omega+1)^2$, in particular whenever
$\underline c^{\,2}\le 8/3$, and then
$\mu=\underline c^{\,2}/\big((1+\bar a)^2+1\big)$: the PL strength is set by
the output bound $\underline c$, attenuated only quadratically by the one-step amplification
of the dynamics, and no quantity grows with $N$.
\end{remark}
 
Theorem~\ref{thm:pl} controls only the bottom of the spectrum of $K$. Paired
with an upper bound on $\lmax(K)$, it yields the conditioning of the lifted
problem, to be contrasted with \eqref{eq:cond}; paired with the descent lemma,
it yields linear convergence of gradient descent to the
\emph{zero-loss set} 
\begin{equation}
    \Mset\eqdef\{w\in\Wset:L_0(w)=0\}
    \label{eq:zeroloss_set}
\end{equation}
of lifted models that
reproduce the data exactly.

\begin{corollary}[Conditioning of the lifted problem]\label{cor:cond}
Under the assumptions of Theorem~\ref{thm:pl}, let $w\in\Wset$,
$\bar c\eqdef\max_{k}\norm{C_k}$ and
$\Lambda_\theta\eqdef\sum_{k=1}^N\norm{B_k}^2$. Then
\begin{equation*}
  \frac{\mu}{2}\le\lmin(K)\le\bar c^{\,2},
  \qquad
  \frac{\Lambda_\theta}{\nth}\le\lmax(K)\le\Lambda_0+\Lambda_\theta,
\end{equation*}
where $\Lambda_0\eqdef 1+(1+\bar a)^{2}+\bar c^{\,2}+\bar a^{2}$ involves only
one-step quantities. Consequently,
$\Lambda_\theta/(\nth\bar c^{\,2})\le\kappa(K)\le2(\Lambda_0+\Lambda_\theta)/\mu$.
\end{corollary}
\begin{proof}See Appendix~\ref{app:pf-cor-cond}.\end{proof}

The curvature of the lifted problem thus grows at most linearly in $N$,
against the geometric growth \eqref{eq:cond} of single shooting, and only
through the parameter term $\Lambda_\theta$, contributed by the rank-$\nth$
block $B_e$, while the floor $\lmin(K)\ge\mu/2$ is unaffected. Moreover, with
$\bar c$ bounded, $\kappa(K)$ is bounded uniformly in $N$ if and only if
$\Lambda_\theta$ is, i.e., if and only if the parameter sensitivities are
square-summable along the trajectory.

\begin{corollary}[Linear convergence]\label{cor:linconv}
Let Assumption~\ref{ass:smooth} hold, let $\ny\le\nx$, and let gradient descent
$w^{t+1}=w^t-\eta\,\nabla L_0(w^t)$ generate iterates with
$\mu(w^t)\ge\underline\mu>0$ for all $t\ge0$. If $\nabla L_0$ is $\ell_0$-Lipschitz on
the convex hull of $\{w^t\}_{t\ge0}$ and $\eta\in(0,1/\ell_0]$, then
\begin{equation}
  L_0(w^t)\le(1-\underline\mu\eta)^{t}\,L_0(w^0),\qquad t=0,1,2,\dots,
  \label{eq:linconv}
\end{equation}
and the iterates converge to a point $w^\infty$ with $L_0(w^\infty)=0$. In
particular, if all iterates lie in a closed set $\Wset$ on which
Assumptions~\ref{ass:submersion} and~\ref{ass:bounded} hold, then
\eqref{eq:linconv} holds with $\underline\mu=\mu$ of \eqref{eq:pl-const}, and
$w^\infty\in\Mset$.
\end{corollary}
\begin{proof}See Appendix~\ref{app:pf-cor-linconv}.\end{proof}

The hypotheses of Corollary~\ref{cor:linconv} involve only the iterates. A
bounded sublevel set $\{w:L_0(w)\le L_0(w^0)\}$, the classical way to confine the iterates,
is not available, since the correction can absorb any change of $\theta$, as
the next section shows. The PL constant is horizon-uniform, whereas the rate
$1-\mu\eta$ is so only if $\ell_0$ is. Near the zero-loss set, $\ell_0$ is set by the
Gauss--Newton part $2\Jr\trans\Jr$ of the Hessian of $L_0$, whose norm
$2\lmax(K)\le2(\Lambda_0+\Lambda_\theta)$ grows, by Corollary~\ref{cor:cond}, at
most linearly in $N$ when the $\norm{B_k}$ are bounded, and not geometrically as
in \eqref{eq:cond}.

This answers (Q1): on the lifted problem, gradient descent reaches a global
minimizer at a linear rate, with a PL constant independent of the horizon,
whenever its iterates remain where Assumptions~\ref{ass:submersion}
and~\ref{ass:bounded} hold. Whether such a minimizer also answers (Q2) is a
different matter, since the global minimizers form the whole zero-loss set
$\Mset$; the next section shows that the condition $L_0(w)=0$ alone does not
determine $\theta$.

\section{Zero loss does not identify the system}
\label{sec:curse}

This section describes the zero-loss set $\Mset$: it is a $C^1$ manifold {of
dimension $p-m=\nth+\nx+N(\nx-\ny)$, by \eqref{eq:margin},} which grows linearly
in $N$ when $\nx>\ny$, and, over the full space, its
projection onto the parameter space
covers every value of $\theta$. We work under exact realizability, to isolate
this geometry from estimation error.

\begin{assumption}[Exact realizability]\label{ass:realiz}
The record is noiseless and generated by \eqref{eq:system}: there exist
$(\theta^\star,x_0^\star)\in\R^{\nth}\times\R^{\nx}$ such that
\begin{equation}
  y_k = g\big(\fk{k}(x_0^\star,\theta^\star)\big),\qquad k=1,\dots,N .
  \label{eq:realiz}
\end{equation}
\end{assumption}

With the true trajectory $z_k^\star\eqdef\fk{k}(x_0^\star,\theta^\star)$, the
true configuration
\begin{equation}
  w^\star\eqdef(\theta^\star,\,x_0^\star,\,z^\star,\,0)
  \label{eq:wstar}
\end{equation}
satisfies $e(w^\star)=0$ by construction and $s(w^\star)=0$ by
\eqref{eq:realiz}; hence $r(w^\star)=0$, with no rank condition involved.

The points of the lifted space used from here on, starting with $w^\star$, follow two conventions. A superscript or subscript on $w$ applies to each of its blocks, e.g., $w^\dagger=(\theta^\dagger,x_0^\dagger,z^\dagger,d^\dagger)$. For a set $\mathcal S\subseteq\R^p$, $\pi_\bullet(\mathcal S)$ denotes its projection onto the blocks named by $\bullet$, e.g., $\pi_\theta(\mathcal S)\eqdef\{\theta:(\theta,x_0,z,d)\in\mathcal S\}$. The \emph{physical} pair is abbreviated as $\chi\eqdef(\theta,x_0)$, with $\Ups(\chi)\eqdef\Ups(\theta,x_0)$, and the same convention applies to it, e.g., $\chi^\star\eqdef(\theta^\star,x_0^\star)$.
 
\begin{proposition}[Geometry of the zero-loss set]\label{prop:manifold}
Under Assumptions~\ref{ass:smooth}--\ref{ass:realiz}, with
$w^\star\in\Wset$, the zero-loss set $\Mset$ \eqref{eq:zeroloss_set} is nonempty, and near each of its
points in $\operatorname{int}\Wset$ it is a $C^1$ embedded submanifold of
dimension
\begin{equation}
  \dim\Mset = p-m,
  \label{eq:dimM}
\end{equation}
given by \eqref{eq:margin}.
If $\Wset$ is open, this holds at every point of $\Mset$.
\end{proposition}
\begin{proof}
Nonemptiness holds because $r(w^\star)=0$ and $w^\star\in\Wset$. By
Theorem~\ref{thm:pl}, $\smin(\Jr(w))\ge\sqrt{\mu/2}>0$ on $\Wset$, so $0$ is a
regular value of the restriction of the $C^1$ map $r$ to
$\operatorname{int}\Wset$, and by the regular value theorem
$r^{-1}(0)\cap\operatorname{int}\Wset$ is a $C^1$ embedded submanifold of
codimension $m$.
\end{proof}
 
A perfect fit is thus achieved by a continuum of models, of dimension at least $\nth+\nx$. Its projection onto the parameter space is as large as possible, as the following proposition shows.
 
\begin{proposition}[Non-identifiability on the zero-loss set] \label{prop:curse}
Under Assumptions~\ref{ass:smooth} and~\ref{ass:realiz}, with the zero-loss set taken over the full
space ($\Wset=\R^p$), $\pi_\theta(\Mset)=\R^{\nth}$: for every $\theta$ there
exists a completion $(x_0,z,d)$ with $(\theta,x_0,z,d)\in\Mset$, so the
condition $L_0(w)=0$ imposes no constraint on $\theta$. For a proper region
$\Wset\subsetneq\R^p$, the same holds for every $\theta$ whose completion lies
in $\Wset$, a set containing a neighborhood of $\theta^\star$ whenever
$w^\star\in\operatorname{int}\Wset$.
\end{proposition}

\begin{proof}
Fix $\theta\in\R^{\nth}$ and let
$d_k(\theta)\eqdef z_k^\star-f(z_{k-1}^\star,\theta)$, $k=1,\dots,N$. Then
$w(\theta)\eqdef(\theta,x_0^\star,z^\star,d(\theta))$ satisfies
$e(w(\theta))=0$ by construction and $s(w(\theta))=s(w^\star)=0$, so
$L_0(w(\theta))=0$ and the parameter block of $w(\theta)$ is $\theta$; hence
$w(\theta)\in\Mset$ whenever $w(\theta)\in\Wset$, in particular always when
$\Wset=\R^p$. The last claim follows because $w(\theta^\star)=w^\star$ and
$\theta\mapsto w(\theta)$ is continuous, $f$ being continuous by
Assumption~\ref{ass:smooth}.
\end{proof}

The construction uses no rank condition: the non-identifiability follows from the freedom of
the correction alone. The feature that yields the PL inequality, a correction
rich enough to absorb any dynamic mismatch, makes a perfect fit uninformative
about the physics, so optimization and identification are distinct problems on
this landscape.

\section{Identifiability and selection}
\label{sec:selection}
Among all models that fit the data perfectly, we select the one whose
correction is smallest. This encodes the physics-informed prior that
$f(\cdot,\theta)$ should explain the data as far as possible, the correction
being confined to genuinely unmodeled effects. Formally, the
\emph{minimal-correction selector} is
\begin{equation}
  w^\dagger \in \argmin_{w\in\Mset}\ \norm{d}^2 .
  \label{eq:selector}
\end{equation}
 
\subsection{Exact recovery under structural identifiability}
\label{subsec:exact}

Identifiability of the physical parameter is a property of the
output map $\Ups$ of \eqref{eq:Upsilon}, i.e., of the model with the correction
set to zero.

\begin{assumption}[Identifiability and excitation at the truth] \label{ass:ident}
The only $\chi\in\pi_\chi(\Wset)$ with $\Ups(\chi)=\Ups(\chi^\star)$ is
$\chi^\star$, and $\partial\Ups(\chi^\star)$ has full column rank $\nth+\nx$.
\end{assumption}

The rank condition is the joint observability and persistency-of-excitation
requirement: the $\theta$-columns of $\partial\Ups$ encode parameter
sensitivity, and the $x_0$-columns the observability of the linearized
trajectory. It is the state-space analogue of the full-column-rank regressor
condition of regression-based identification.

Under this assumption, minimal correction removes the ambiguity of
Proposition~\ref{prop:curse}: among all exact fits, the truth is the only one
that needs no correction.
\begin{theorem}[Exact recovery by the selector]\label{thm:exact}
Under Assumptions~\ref{ass:realiz} and~\ref{ass:ident}, with
$w^\star\in\Wset$, the true configuration is the unique global minimizer in
\eqref{eq:selector}, i.e., $\argmin_{w\in\Mset}\norm{d}^2=\{w^\star\}$, so
that $w^\dagger=w^\star$ and, in particular, $\theta^\dagger=\theta^\star$.
\end{theorem}
\begin{proof}
Since $w^\star\in\Mset$ and $d^\star=0$, $w^\star$ is a minimizer and every
minimizer $w^\dagger$ satisfies $d^\dagger=0$. With $d^\dagger=0$ and
$z_0^\dagger=x_0^\dagger$, the equations $e(w^\dagger)=0$ unfold to
$z_k^\dagger=\fk{k}(x_0^\dagger,\theta^\dagger)$, and $s(w^\dagger)=0$ then
gives $\Ups(\chi^\dagger)=\big(g(z_k^\dagger)\big)_{k=1}^N=y=\Ups(\chi^\star)$
by \eqref{eq:realiz}. Since $w^\dagger\in\Wset$, $\chi^\dagger\in\pi_\chi(\Wset)$,
so Assumption~\ref{ass:ident} gives $\chi^\dagger=\chi^\star$, then
$z^\dagger=z^\star$, and $w^\dagger=w^\star$.
As $w^\dagger$ was an arbitrary minimizer, $w^\star$ is the unique~one.
\end{proof}
Exact recovery thus only requires identifiability of the physical model itself.
Theorem~\ref{thm:exact} is, however, a statement about the selector, which is defined
through the implicit constraint $w\in\Mset$; the next subsection makes it
computable.

\subsection{Selection by penalization}
\label{subsec:regpath}

The selector \eqref{eq:selector} cannot be optimized directly, since its
feasible set $\Mset$ is known only implicitly through $r(w)=0$. The computable
surrogate is $L_\gamma$ \eqref{eq:Lgam}, which trades the constraint $w\in\Mset$ for the fit
term and, under realizability, realizes the selector at every positive weight.

\begin{theorem}[Selection by penalization]\label{thm:regpath}
Let Assumption~\ref{ass:smooth} hold and let $\Wset$ be compact. Then $L_0$
and, for every $\gamma>0$, $L_\gamma$ are continuous and attain their minima
over $\Wset$; let $w_\gamma$ be a
minimizer of $L_\gamma$ over $\Wset$.
\begin{enumerate}
    \item[(i)] If Assumption~\ref{ass:realiz} holds and $w^\star\in\Wset$, then $w_\gamma\in\Mset$ and $d_\gamma=0$ for every $\gamma>0$; if, moreover, Assumption~\ref{ass:ident} holds, then $w_\gamma=w^\star$.
    \item[(ii)] { If $\Mset\neq\emptyset$, every accumulation point of
$\{w_\gamma\}$ as $\gamma\downarrow0$ solves the selector \eqref{eq:selector}.}
\end{enumerate}
\end{theorem}
 
\begin{proof}
(i) Since $w^\star\in\Wset$, $r(w^\star)=0$ and $d^\star=0$,
$0\le L_\gamma(w_\gamma)\le L_\gamma(w^\star)=0$, so both nonnegative terms of
$L_\gamma(w_\gamma)$ vanish: $w_\gamma\in\Mset$ and $d_\gamma=0$. Hence
$w_\gamma$ minimizes $\norm{d}^2$ over $\Mset$, and Theorem~\ref{thm:exact}
gives $w_\gamma=w^\star$ under Assumption~\ref{ass:ident}.
The proof of (ii) is in Appendix~\ref{app:pf-thm-regpath}.
\end{proof}
 
Part~(i) is the operative statement under exact realizability: the
\emph{global} minimizer of $L_\gamma$ over $\Wset$ is the true configuration
at every positive weight, with no vanishing-$\gamma$ limit. This is a property
of the estimator, not of an algorithm, which must still reach that minimizer.
Part~(ii) requires no realizability, only an exact fit in $\Wset$: the
accumulation points then solve the selector \eqref{eq:selector}, and the size
of the selected correction quantifies the unmodeled part.

Whether gradient descent can reach that minimizer depends on the landscape of
$L_\gamma$. Since $L_\gamma$ differs from $L_0$ only by the term
$\gamma\norm{d}^2$, the PL inequality of Theorem~\ref{thm:pl} survives the
penalty up to an error of order $\gamma$, as the following proposition shows.
\begin{proposition}[Inexact global PL for the penalized loss]\label{prop:inexactPL}
Let Assumptions~\ref{ass:smooth}--\ref{ass:realiz} hold on a compact $\Wset$
with $w^\star\in\Wset$, let $\mu$ be the constant of Theorem~\ref{thm:pl}, and
set $M_d\eqdef\sup_{w\in\Wset}\norm{d}$. Then, for every $\gamma>0$ and every
$w\in\Wset$,
\begin{equation}
  \tfrac12\norm{\nabla L_\gamma(w)}^2\ge\tfrac{\mu}{2}L_\gamma(w)-\Xi_\gamma,
  \qquad
  \Xi_\gamma\eqdef\tfrac{\mu}{2}\gamma M_d^2+2\gamma^2M_d^2,
  \label{eq:inexactPL}
\end{equation}
and $\min_{\Wset}L_\gamma=L_\gamma(w^\star)=0$.
\end{proposition}
\begin{proof}See Appendix~\ref{app:pf-prop-inexactPL}.\end{proof}

As in Corollary~\ref{cor:linconv}, the descent lemma turns
\eqref{eq:inexactPL} into a convergence statement, now up to a floor of
order $\gamma$.

\begin{corollary}[Global descent on the penalized loss] \label{cor:inexactPL}
Under the hypotheses of Proposition~\ref{prop:inexactPL}, let $\ell_\Wset\ge2$ bound the Lipschitz modulus of $\nabla L_0$ on the convex
hull of $\Wset$ (no restriction, since $\partial_d^2L_0=2\bI$), and
let $w^{t+1}=w^t-\eta\nabla L_\gamma(w^t)$ with $\eta\le1/(\ell_\Wset+2\gamma)$ and
$w^t\in\Wset$ for all $t$. Then
\begin{align*}
  L_\gamma(w^t)&\le\Big(1-\frac{\eta\mu}{2}\Big)^{t}L_\gamma(w^0)+\frac{2\Xi_\gamma}{\mu},\\
  \limsup_{t\to\infty}L_0(w^t)&\le\frac{2\Xi_\gamma}{\mu}
  =\gamma M_d^2+\frac{4\gamma^2M_d^2}{\mu}.
\end{align*}
\end{corollary}
\begin{proof}See Appendix~\ref{app:pf-cor-inexactPL}.\end{proof}

The fit guarantee of Corollary~\ref{cor:linconv} thus survives penalization up
to an $O(\gamma)$ floor. The bound does not discriminate among zero-loss
models, however: every $w\in\Mset$ satisfies
$L_\gamma(w)=\gamma\norm d^2\le\gamma M_d^2\le2\Xi_\gamma/\mu$, so the sublevel
set $\{L_\gamma\le2\Xi_\gamma/\mu\}$, to which the bound confines the iterates
asymptotically, contains all of $\Mset$. 

An exact PL inequality for $L_\gamma$ would discriminate, since it would drive
the iterates to $w^\star$, the only zero of $L_\gamma$
(Theorem~\ref{thm:exact}), but no such inequality holds in general, for two reasons.
First, Lemma~\ref{lem:pl} does not apply: writing
$L_\gamma=\norm{(e,s,\sqrt\gamma\,d)}^2$ adds one residual row per correction
variable, and the margin \eqref{eq:margin} drops to $\nth+\nx-N\ny$, negative
for $N>(\nth+\nx)/\ny$. Second, whenever the selection objective
$\norm{d}^2$ has a nonglobal strict local minimizer on $\Mset$, as it may for a
nonlinear, partially observed system, $L_\gamma$ has spurious local
minimizers, as the next proposition shows.

\begin{proposition}[Spurious local minimizers of the penalized loss]\label{prop:spurious}
Let Assumptions~\ref{ass:smooth},~\ref{ass:realiz} and~\ref{ass:ident} hold
with $w^\star\in\Wset$, and let
$\bar w\in(\Mset\cap\operatorname{int}\Wset)\setminus\{w^\star\}$, with
correction $\bar d$, be a strict local minimizer of $\norm{d}^2$ on $\Mset$.
Then every neighborhood of $\bar w$ contains, for all sufficiently small
$\gamma>0$, a local minimizer $w'_\gamma$ of $L_\gamma$ with
$L_\gamma(w'_\gamma)>0=\min_{\Wset}L_\gamma$. In particular, $L_\gamma$ satisfies no
PL inequality on a set containing $w'_\gamma$.
\end{proposition}
\begin{proof}See Appendix~\ref{app:pf-prop-spurious}.\end{proof}

Proposition~\ref{prop:spurious} thus separates the two tasks. Fitting the
data is globally benign: $L_0$ satisfies a PL inequality, and gradient descent
reaches $\Mset$ under the conditions of Corollary~\ref{cor:linconv}, also in
the presence of a small penalty, up to an $O(\gamma)$ floor
(Corollary~\ref{cor:inexactPL}). Selecting $w^\star$ on $\Mset$ is not:
$L_\gamma$ inherits the nonglobal minimizers of the selection objective
$\norm{d}^2$ on $\Mset$, so no global guarantee for selection is available in
general. We therefore minimize $L_0$, which satisfies the PL inequality, to reach $\Mset$, and use
the penalty only for selection, where the guarantees are local.

\subsection{Homotopy continuation in the weight}
\label{subsec:homotopy}
These local guarantees rest on the curvature of $L_\gamma$ at $w^\star$, which
the penalty makes strictly positive. This curvature yields strong convexity on
a ball whose radius is proportional to the weight, and continuation in the
weight carries the iterate through a sequence of such balls, from a large weight
down to the target one.

\begin{proposition}[Curvature of the penalized loss at the truth]\label{prop:localPL}
Let Assumptions~\ref{ass:smooth}--\ref{ass:ident} hold with $w^\star\in\Wset$.
For $\gamma>0$, let $r_\gamma\eqdef(r,\sqrt\gamma\,d)$, so that
$L_\gamma=\norm{r_\gamma}^2$, and let $\Jr_\gamma\eqdef\partial r_\gamma$. Then
\begin{enumerate}
\item[(i)] $\Jr_\gamma(w^\star)$ has full column rank, i.e., $\ker\Jr_\gamma(w^\star)=\{0\}$;
\item[(ii)] with
$\rho_\star\eqdef\min\{\norm{v_d}^2:v=(v_\theta,v_{x_0},v_z,v_d)\in\ker\Jr(w^\star),\ \norm v=1\}>0$,
for every $\gamma\in(0,\mu/2]$,
\[
  \gamma\rho_\star\big(1-2\gamma/\mu\big)
  \le\lmin\big(\Jr_\gamma(w^\star)\trans\Jr_\gamma(w^\star)\big)
  \le\gamma\rho_\star ;
\]
    \item[(iii)]  $\rho_\star\le\Omega\,\sigma_\star^2/\underline c^{\,2}$, where
$\sigma_\star\eqdef\smin(\partial\Ups(\chi^\star))$.
\end{enumerate}
\end{proposition}
\begin{proof}See Appendix~\ref{app:pf-prop-localPL}.\end{proof}

At $w^\star$ the penalized residual vanishes, so, where $f$ and $g$ are twice
differentiable, the Hessian of $L_\gamma$ at $w^\star$ is
$2\Jr_\gamma(w^\star)\trans\Jr_\gamma(w^\star)$, which is positive definite for
every $\gamma>0$ by part~(i). Hence $w^\star$ is a nondegenerate local
minimizer of $L_\gamma$ at every weight. Unlike the PL inequality of
Theorem~\ref{thm:pl}, this property is local: there, the correction gives
$\Jr$ full row rank on all of $\Wset$, whereas here identifiability gives
$\Jr_\gamma$ full column rank only at $w^\star$.

By part~(ii), the curvature at $w^\star$ is proportional to $\gamma$, so it is
smallest at the small weights, at which $L_\gamma$ is closest to the PL
landscape of $L_0$ (Corollary~\ref{cor:inexactPL}). By part~(iii), it is also
limited by identifiability: a small singular value $\sigma_\star$ forces a
small $\rho_\star$, hence a small curvature at every weight. The following
lemma shows that $L_\gamma$ is strongly convex on a ball around $w^\star$ whose
radius is also proportional to $\gamma$.

\begin{lemma}[Local strong convexity of the penalized loss]\label{lem:scLgamma}
Let the assumptions of Proposition~\ref{prop:localPL} hold, and let $f$ and
$g$ be twice continuously differentiable on neighborhoods of the points
$(z_{k-1}^\star,\theta^\star)$ and $z_k^\star$, $k=1,\dots,N$, respectively.
Then there exist a closed ball $\mathcal U$ centered at $w^\star$ and
$\ell_H>0$ such that
$\nabla^2L_0(w)\succeq\nabla^2L_0(w^\star)-\ell_H\norm{w-w^\star}\bI$ for all
$w\in\mathcal U$. Moreover, let $\gamma_{\max}\in(0,\mu/2)$ bound the weights
considered, and let 
$\sigma_0\eqdef\rho_\star(1-2\gamma_{\max}/\mu)$,
$\varrho\eqdef\sigma_0/\ell_H$ and
$\mathcal B_\gamma\eqdef\{w:\norm{w-w^\star}\le\varrho\gamma\}$, with
$\gamma_{\max}$ small enough that $\mathcal B_{\gamma_{\max}}\subseteq\mathcal U$
(possible, since $\varrho\gamma_{\max}\to0$ as $\gamma_{\max}\downarrow0$).
Then, for every $\gamma\in(0,\gamma_{\max}]$, $L_\gamma$ is
$\sigma_0\gamma$-strongly convex and $\ell_B$-smooth on $\mathcal B_\gamma$,
with $\ell_B\eqdef\max_{\mathcal U}\norm{\nabla^2L_0}+2\gamma_{\max}$, and
$w^\star=\argmin_{\mathcal B_\gamma}L_\gamma$.
\end{lemma}
\begin{proof}See Appendix~\ref{app:pf-lem-scLgamma}.\end{proof}

Both the strong-convexity modulus $\sigma_0\gamma$ and the radius $\varrho\gamma$
are proportional to $\gamma$: the basin around $w^\star$ shrinks with the
penalty. Under exact realizability, $w^\star$ minimizes $L_\gamma$ for every
$\gamma>0$ (Theorem~\ref{thm:regpath}(i)), so a single solve at $\gamma_{\max}$,
where the basin is largest, suffices. Without realizability, a small $\gamma$ is needed.

The minimizer of $L_\gamma$ then moves with $\gamma$, and only its limit as
$\gamma\downarrow0$ solves the selector (Theorem~\ref{thm:regpath}(ii)). At a small
weight, however, the basin of Lemma~\ref{lem:scLgamma} is small, so a single
solve must start close to the minimizer. Continuation relaxes this
requirement: it starts at a large weight, where the basin is largest, and
decreases the weight in stages, each started from the output of the previous
one and ending inside the basin of the next. Its analysis
is carried out under exact realizability, where the minimizer is $w^\star$ at
every stage; the case in which the minimizer moves with $\gamma$ is left for future work.

Fix a target weight $\gamma_\star>0$, at which the estimate is returned, an
initial weight $\gamma_0\ge\gamma_\star$ and a number $T\ge1$ of weight
reductions, and consider the geometric schedule $\gamma_j\eqdef\gamma_0\tau^j$,
$j=0,\dots,T$, with ratio $\tau\eqdef(\gamma_\star/\gamma_0)^{1/T}\in(0,1]$,
which ends at $\gamma_T=\gamma_\star$. Starting from $w^{(0)}$, typically the output of gradient
descent on $L_0$ (Corollary~\ref{cor:linconv}), each of the $T+1$ stages
runs $n_j$ gradient steps of size $\eta$ on
$L_{\gamma_j}$ from the output of the previous one, as summarized in
Algorithm~\ref{alg:homotopy}. The following proposition gives admissible
$\eta$ and $n_j$. In practice, any local descent method can replace the
gradient steps, e.g., Levenberg--Marquardt
iterations~\cite{levenberg1944LeastSquares,marquardt1963LeastSquares}, since
$L_{\gamma_j}=\norm{r_{\gamma_j}}^2$ is a least-squares cost.
\begin{algorithm}[t]
\caption{Homotopy continuation for selection}\label{alg:homotopy}
\begin{algorithmic}[1]
\Require data $y$; start $w^{(0)}$;
         weights $\gamma_0\ge\gamma_\star$; reductions $T$; step size $\eta$; steps per stage $n_0,\dots,n_T$
\State (optional) $w^{(0)}\gets$ gradient descent on $L_0$ from $w^{(0)}$
       \Comment{reach $\Mset$, Cor.~\ref{cor:linconv}}
\State $\gamma_j\gets\gamma_0(\gamma_\star/\gamma_0)^{j/T}$, $j=0,\dots,T$
\For{$j=0$ to $T$}
\State $w^{(j+1)}\gets w^{(j)}$ \Comment{warm start}
\For{$i=1$ to $n_j$}
\State $w^{(j+1)}\gets w^{(j+1)}-\eta\nabla L_{\gamma_j}(w^{(j+1)})$
\EndFor
\EndFor
\State \Return $w^{(T+1)}$
\end{algorithmic}
\end{algorithm}

Continuation succeeds if each stage ends inside the basin of the next. Since
the radii $\varrho\gamma_j$ shrink by the factor $\tau$ from one weight to the
next, it suffices that each stage reduce the distance to $w^\star$ by that
factor, which gradient steps on the $\sigma_0\gamma_j$-strongly convex
$L_{\gamma_j}$ achieve in a number of iterations proportional to
$1/\gamma_j$, as the following proposition shows.
\begin{proposition}[Convergence of the continuation]\label{prop:homotopy}
Adopt the hypotheses and notation of Lemma~\ref{lem:scLgamma}, and let
$\gamma_\star<\gamma_0\le\gamma_{\max}$, so that $\tau<1$, and run
Algorithm~\ref{alg:homotopy} with $\eta\le1/\ell_B$,
$n_j\ge\ln(1/\tau)/(\eta\sigma_0\gamma_j)$ for $j=0,\dots,T-1$, and
$n_T\ge0$. If $w^{(0)}\in\mathcal B_{\gamma_0}$, then
$w^{(j)}\in\mathcal B_{\gamma_j}$ for $j=0,\dots,T$, and
$\norm{w^{(T+1)}-w^\star}\le(1-\eta\sigma_0\gamma_\star)^{n_T}\varrho\gamma_\star$;
in particular, $\norm{\theta^{(T+1)}-\theta^\star}$ obeys the same bound.
\end{proposition}
\begin{proof}See Appendix~\ref{app:pf-prop-homotopy}.\end{proof}

Proposition~\ref{prop:homotopy} quantifies the advantage of
Algorithm~\ref{alg:homotopy} over a single solve at the target weight. Since
$L_\gamma$ is minimized by $w^\star$ at every weight
(Theorem~\ref{thm:regpath}(i)), continuation does not follow a moving target:
it carries the iterate through basins that shrink with the weight. A single
solve at $\gamma_\star$ is guaranteed to converge from distance
$\varrho\gamma_\star$ of $w^\star$, continuation from distance
$\varrho\gamma_0$, larger by the factor $\gamma_0/\gamma_\star$. The price is
small: since $\gamma_j=\gamma_0\tau^j$ and $\ln(1/\tau)\le1/\tau-1$, with the
smallest admissible $n_j$ the first
$T$ stages take at most $1/(\eta\sigma_0\gamma_\star)+T$ gradient steps in
total, independently of $\gamma_0$. 

\begin{remark}[Scope of continuation]\label{rem:homotopyscope}
Proposition~\ref{prop:homotopy} has one hypothesis that the preceding results
do not guarantee: the start $w^{(0)}$ must lie in $\mathcal B_{\gamma_0}$.
Continuation weakens this requirement without removing it: a single solve at
the target weight must start in $\mathcal B_{\gamma_\star}$, of radius
$\varrho\gamma_\star$, whereas continuation needs a start in
$\mathcal B_{\gamma_0}$, of radius $\varrho\gamma_0$, largest for
$\gamma_0=\gamma_{\max}$. The fit phase does not supply such a start: it
reaches $\Mset$, and Corollary~\ref{cor:inexactPL} confines the iterates to a
sublevel set that contains all of $\Mset$, while $\mathcal B_{\gamma_0}$ is a ball around $w^\star$. This requirement disappears when $r$ is affine in
$w$, since $L_\gamma$ is then a quadratic with constant Hessian
$2\Jr_\gamma\trans\Jr_\gamma$, positive definite by
Proposition~\ref{prop:localPL}(i), so that $\ell_H$ can be taken arbitrarily small and $\mathcal B_{\gamma_0}$ arbitrarily large. It cannot be guaranteed in general, because
$L_\gamma$ may have local minimizers other than $w^\star$
(Proposition~\ref{prop:spurious}), and a descent method started in the
basin of one of them can remain there. The guarantee of
Proposition~\ref{prop:homotopy} is therefore local by necessity.
\end{remark}

\subsection{Robustness to noise and unmodeled dynamics}
\label{subsec:robust}

We now relax Assumption~\ref{ass:realiz}. Because the correction enters the
state recursion freely, any output record whose samples lie in the range of
$g$ is interpolated by a lifted configuration: for any $\theta$ and $x_0$,
choose $z_k$ with $g(z_k)=y_k$ and set $d_k\eqdef z_k-f(z_{k-1},\theta)$.
The range of $g$ is all of $\R^{\ny}$ for affine outputs of full row rank. In
general, $g$ is a submersion at the lifted states of $\Wset$
(Assumption~\ref{ass:submersion}), hence open there, so every record close
enough to one attainable by the model within $\Wset$ still has its samples in
the range of $g$.
We therefore adopt as data model the existence of a true configuration in
$\Wset$,
\begin{equation}
\begin{aligned}
  w^\star&=(\theta^\star,x_0^\star,z^\star,d^\star)\in\Mset,\\
  z_k^\star&=f(z_{k-1}^\star,\theta^\star)+d_k^\star, \quad y_k=g(z_k^\star),
\end{aligned}
\label{eq:truecorrected}
\end{equation}
which reduces to \eqref{eq:wstar} when $d^\star=0$. Its correction $d^\star$
aggregates the unmodeled state-transition mismatch and the state perturbation
induced by the measurement noise, and $\norm{d^\star}$ measures the departure
from exact realizability. The pair $(z^\star,d^\star)$ is not unique, since any
lifted trajectory through the measured outputs defines one, and the results
below hold for every such choice, in particular for one of minimal
$\norm{d^\star}$. We take $\Wset$ compact, so that $\Mset$ is
compact and the selector \eqref{eq:selector} exists. Recovery is no longer
exact, and the bias is controlled by $\norm{d^\star}$, by the output
sensitivity to the correction, and by the conditioning of $\Ups$, which we make
quantitative.

\begin{assumption}[Quantitative identifiability]\label{ass:qualident}
There exists $\underline\upsilon>0$ such that
\begin{equation}
  \norm{\Ups(\chi)-\Ups(\chi^\star)}\ge\underline\upsilon\,\norm{\chi-\chi^\star},
  \qquad \forall\,\chi\in\pi_\chi(\Wset).
  \label{eq:lowermodulus}
\end{equation}
\end{assumption}

Condition \eqref{eq:lowermodulus} strengthens the first part of
Assumption~\ref{ass:ident} from uniqueness to stability: output errors bound
parameter errors. When
$\chi^\star$ is interior to $\pi_\chi(\Wset)$, letting $\chi\to\chi^\star$ in
\eqref{eq:lowermodulus} gives $\smin(\partial\Ups(\chi^\star))\ge\underline\upsilon$,
so \eqref{eq:lowermodulus} implies the rank condition of
Assumption~\ref{ass:ident} and $\underline\upsilon\le\sigma_\star$, with equality
admissible when $\Ups$ is affine.

The effect of a correction on the output trajectory is controlled by
sensitivity bounds along the segments joining the lifted and simulated
trajectories. Let $\Wset^+\subseteq\R^{\nx}\times\R^{\nth}$ be the set of
pairs $(x,\theta)$ such that, for some $w=(\theta,x_0,z,d)\in\Wset$ and some
$k\in\{0,\dots,N\}$, $x$ lies on the segment joining $z_k$ to
$\fk{k}(x_0,\theta)$; it is compact, as a continuous image of a compact set.
Set
\begin{equation}
  \bar a_+\eqdef\sup_{\Wset^+}\norm{\ddx f},\quad
  \ell_g\eqdef\sup_{\Wset^+}\norm{\ddx g},\quad
  S_N\eqdef\sum_{i=0}^{N-1}\bar a_+^{\,i}.
  \label{eq:SN}
\end{equation}
The constant $\bar a_+$ bounds $a(w)$ of \eqref{eq:scalars} on $\Wset$ and
extends it to the joining segments. The horizon factor satisfies
$S_N=(\bar a_+^{\,N}-1)/(\bar a_+-1)$ for $\bar a_+\neq1$, $S_N=N$ for
$\bar a_+=1$, and $S_N\le(1-\bar a_+)^{-1}$ for $\bar a_+<1$.

\begin{lemma}[Output sensitivity to the correction]\label{lem:outsens}
Under Assumption~\ref{ass:smooth}, for every $w=(\theta,x_0,z,d)\in\Wset$,
with $\chi=(\theta,x_0)$,
\begin{equation}
  \norm{\Ups(\chi)-y}\le\ell_g\,S_N\big(\norm{d}+\norm{e(w)}\big)+\norm{s(w)};
  \label{eq:outsens}
\end{equation}
in particular, $\norm{\Ups(\chi)-y}\le\ell_gS_N\norm{d}$ for every $w\in\Mset$.
\end{lemma}
\begin{proof}See Appendix~\ref{app:pf-lem-outsens}.\end{proof}

\begin{theorem}[Robust recovery]\label{thm:robust}
Under Assumptions~\ref{ass:smooth} and~\ref{ass:qualident}, with $\Wset$ compact, let the data be
generated as in \eqref{eq:truecorrected} and let
$w^\dagger=(\theta^\dagger,x_0^\dagger,z^\dagger,d^\dagger)$ be any solution of
\eqref{eq:selector}. Then
\begin{equation}
  \norm{\theta^\dagger-\theta^\star}
   \le \norm{\chi^\dagger-\chi^\star}
   \le \frac{2\,\ell_g\,S_N}{\underline\upsilon}\,\norm{d^\star}.
  \label{eq:robustbound}
\end{equation}
\end{theorem}
\begin{proof}
Both $w^\star$ and $w^\dagger$ lie in $\Mset$, and $w^\dagger$ minimizes
$\norm{d}^2$ over $\Mset$, so $\norm{d^\dagger}\le\norm{d^\star}$.
Lemma~\ref{lem:outsens}, applied to both, bounds $\norm{\Ups(\chi^\dagger)-y}$
and $\norm{\Ups(\chi^\star)-y}$ by $\ell_gS_N\norm{d^\star}$, and the triangle
inequality gives
$\norm{\Ups(\chi^\dagger)-\Ups(\chi^\star)}\le2\ell_gS_N\norm{d^\star}$. The
lower modulus \eqref{eq:lowermodulus} bounds the left-hand side from below by
$\underline\upsilon\norm{\chi^\dagger-\chi^\star}$, which yields the second
inequality in \eqref{eq:robustbound}; the first is immediate.
\end{proof}

The bound collapses to Theorem~\ref{thm:exact} as $d^\star\to0$, so the
estimator interpolates continuously between exact recovery and a bias at most
linear in the unmodeled effects, amplified by the inverse stability modulus
$\underline\upsilon^{-1}$ of the parameter-to-output map. The horizon factor
$S_N$ is bounded uniformly in $N$ under the incremental contraction
$\bar a_+<1$ and grows with $N$ otherwise. This is the only place where
stability enters the analysis, and it enters the \emph{identification}
constant alone: the PL constant of Theorem~\ref{thm:pl} is horizon-uniform
whatever the value of $\bar a$.

\begin{corollary}[Robust recovery at a positive weight]\label{cor:robustgamma}
Under the assumptions of Theorem~\ref{thm:robust}, let $\gamma>0$ and let
$w_\gamma$ be a minimizer of $L_\gamma$ over $\Wset$.
Then
\begin{equation}
  \norm{\theta_\gamma-\theta^\star}\le\norm{\chi_\gamma-\chi^\star}
  \le\frac{2\,\ell_gS_N+(\ell_gS_N+1)\sqrt\gamma}{\underline\upsilon}\,\norm{d^\star}.
  \label{eq:robustgamma}
\end{equation}
\end{corollary}
\begin{proof}
Since $w^\star\in\Mset$, optimality of $w_\gamma$ gives
$L_0(w_\gamma)+\gamma\norm{d_\gamma}^2\le L_\gamma(w^\star)=\gamma\norm{d^\star}^2$,
hence $\norm{d_\gamma}\le\norm{d^\star}$ and
$\norm{e(w_\gamma)},\norm{s(w_\gamma)}\le\norm{r(w_\gamma)}\le\sqrt\gamma\,\norm{d^\star}$.
Lemma~\ref{lem:outsens} then bounds $\norm{\Ups(\chi_\gamma)-y}$ by
$\big(\ell_gS_N+(\ell_gS_N+1)\sqrt\gamma\big)\norm{d^\star}$ and
$\norm{\Ups(\chi^\star)-y}$ by $\ell_gS_N\norm{d^\star}$, and the triangle
inequality and \eqref{eq:lowermodulus} conclude as in the proof of
Theorem~\ref{thm:robust}.
\end{proof}

As $\gamma\downarrow0$, the bound \eqref{eq:robustgamma} tends to
\eqref{eq:robustbound}; at a positive target weight, the additional error is
of order $\sqrt\gamma\,\norm{d^\star}$, which quantifies the role of a small
target weight. Like Theorem~\ref{thm:regpath}(i), the corollary concerns the
global minimizer of $L_\gamma$, a property of the estimator; computing it is
the task of Algorithm~\ref{alg:homotopy}.

\begin{remark}[Local Jacobian refinement] \label{rem:jacrefine}
The modulus \eqref{eq:lowermodulus} may be replaced by the smallest singular value
$\sigma_\star>0$ of Proposition~\ref{prop:localPL}(iii). Let
$\pi_\chi(\Wset)$ be convex and $\partial\Ups$ be $\ell_\Ups$-Lipschitz on
it. For $\chi\in\pi_\chi(\Wset)$ with
$\norm{\chi-\chi^\star}\le\sigma_\star/\ell_\Ups$, Taylor's theorem gives
$\norm{\Ups(\chi)-\Ups(\chi^\star)}\ge(\sigma_\star-\tfrac{\ell_\Ups}{2}
\norm{\chi-\chi^\star})\norm{\chi-\chi^\star}\ge\tfrac{\sigma_\star}{2}
\norm{\chi-\chi^\star}$. If the selected estimate $\chi^\dagger$ lies in this
ball, which by \eqref{eq:robustbound} holds whenever
$\norm{d^\star}\le\underline\upsilon\,\sigma_\star/(2\ell_gS_N\ell_\Ups)$,
the proof of Theorem~\ref{thm:robust} therefore yields
\begin{equation}
  \norm{\theta^\dagger-\theta^\star}\le
  \frac{4\,\ell_g\,S_N}{\sigma_\star}\,\norm{d^\star}.
  \label{eq:localbound}
\end{equation}
Since $\underline\upsilon\le\sigma_\star$ whenever $\chi^\star$ is interior,
\eqref{eq:localbound} is never more than twice \eqref{eq:robustbound}, and it
is much sharper when $\underline\upsilon\ll\sigma_\star$.
\end{remark}

Theorem~\ref{thm:regpath}(i), Corollary~\ref{cor:robustgamma} and
Proposition~\ref{prop:homotopy} guide the choice of the weights of
Algorithm~\ref{alg:homotopy}. The target weight trades accuracy against
conditioning. Under exact realizability, $w^\star$ minimizes $L_{\gamma_\star}$ at
every $\gamma_\star>0$. Otherwise, the bound \eqref{eq:robustgamma} exceeds
\eqref{eq:robustbound} by the factor $1+(1+1/(\ell_gS_N))\sqrt{\gamma_\star}/2$, at
most $1+\sqrt{\gamma_\star}$ when $\ell_gS_N\ge1$, so $\gamma_\star$ need only be
small enough to make $\sqrt{\gamma_\star}$ negligible against one. Below that, the
bound improves little, whereas the conditioning worsens: under exact
realizability, the curvature of $L_{\gamma_\star}$ at $w^\star$ is of order
$\gamma_\star\rho_\star$ (Proposition~\ref{prop:localPL}(ii)), and continuation takes
about $1/(\eta\sigma_0\gamma_\star)$ gradient steps.

When $\gamma_\star<\gamma_0$, continuation enlarges the guaranteed start
region: a solve at $\gamma_\star$ converges from distance $\varrho\gamma_\star$
of $w^\star$, the continuation from $\varrho\gamma_0$. With
$\gamma_{\max}=\gamma_0$, the radius
$\varrho\gamma_0=\rho_\star\gamma_0(1-2\gamma_0/\mu)/\ell_H$ is largest at
$\gamma_0=\mu/4$, which never exceeds $1/3$, since $\mu\le4/3$ by
\eqref{eq:pl-const}; as $\mu$ is in general unknown, $\mu(w)/4$ at the output
$w$ of the fit phase, which involves only the one-step quantities $a(w)$ and
$c(w)$, offers an estimate, optimistic since $\mu(w)\ge\mu$.
The number $T$ of reductions affects the cost only additively,
since the first $T$ stages take at most $1/(\eta\sigma_0\gamma_\star)+T$ steps.

\section{Numerical results}
\label{sec:numerics}

This section tests the results of
Sections~\ref{sec:geometry}--\ref{sec:selection} on simulated benchmarks.

\subsection{Benchmarks and implementation}
\label{subsec:num-setup}
The experiments use three systems: a synthetic one, whose constants in
Theorem~\ref{thm:pl} and Corollary~\ref{cor:cond} can be set independently,
the Duffing oscillator, in four variants, and the Lorenz system.

The \emph{synthetic benchmark} has $\nx=2$, $\ny=1$, $\nth=2$, and
\begin{equation}
\begin{aligned}
  x_{k+1}&=a_0\,\mathrm{R}(\phi)\,x_k+\big(0.05\,\theta_1\tanh x_{k,1}
          +b_0\,\theta_2\big)\begin{bmatrix}0,&1\end{bmatrix}\trans,\\
  y_k&=c_0\,x_{k,2},
\end{aligned}
\label{eq:num-synth}
\end{equation}
where $x_{k,i}$ is the $i$-th component of $x_k$, $\mathrm{R}(\phi)$ is the
planar rotation by $\phi=0.3$, $\theta^\star=(1,0.5)$ and
$x_0^\star=(0.4,0.2)$. In \eqref{eq:scalars}, the gain $a_0$ sets $a(w)$ up
to $0.05|\theta_1|$ and $c_0$ sets $c(w)=c_0$. The gain $b_0$ enters only
the sensitivity to $\theta_2$: the second column of $B_k$ in \eqref{eq:sens}
is $b_0\begin{bmatrix}0,&1\end{bmatrix}\trans$, so $b_0$ leaves $a(w)$ and
$c(w)$ unchanged and changes only $\Lambda_\theta$ in
Corollary~\ref{cor:cond}. Unless stated otherwise, $b_0=0.05$ and $c_0=1$.

The \emph{Duffing benchmark} is the free-decay oscillator with state
$x=(q,v)$, discretized by the semi-implicit Euler step
$v_{k+1}=v_k+h(-\delta v_k-\alpha q_k-\beta q_k^3)$, $q_{k+1}=q_k+h\,v_{k+1}$,
$h=0.05$, with output $y_k=q_k$, $\theta=(\delta,\alpha,\beta)$,
$\theta^\star=(0.2,1,0.5)$ and $x_0^\star=(1.5,0)$; its output is a
coordinate projection, so Assumption~\ref{ass:submersion} holds with
$\underline c=1$ on all of $\R^p$. The \emph{lightly damped} variant has
$\theta^\star=(0.05,1,0.5)$. 
Two further variants of the Duffing benchmark, with $\delta=0.2$ known, keep
its dynamics and change only how $\theta$ enters them. The
\emph{reparameterized benchmark} sets
$(\alpha,\beta)=(\theta_1^3-3\theta_1+\theta_2,\ 0.5+0.2\,\theta_1)$, with
$\theta^\star=(0,1)$. This map is a diffeomorphism of $\R^2$, so the model set
and the true $(\alpha,\beta)=(1,0.5)$ are unchanged, but $\theta$ enters $f$
through a cubic instead of affinely. The \emph{fold benchmark} fixes
$\beta=0.5$ and sets $\alpha(\theta)=\theta^3-3\theta-2$ for scalar $\theta$.
This map is not injective: it folds at $\theta=\pm1$, where
$\partial\alpha(\theta)=3\theta^2-3$ vanishes. The truth
$\theta^\star\approx2.104$ is the only real root of $\alpha(\theta)=1$ and
$\partial\alpha(\theta^\star)\neq0$, which give the uniqueness and the rank
condition of Assumption~\ref{ass:ident}.

The \emph{Lorenz benchmark} is the Lorenz-63 system discretized by the
explicit Euler step,
$x_{k+1}=x_k+h\,\big(\sigma(x_{k,2}-x_{k,1}),\ x_{k,1}(\rho-x_{k,3})-x_{k,2},\
x_{k,1}x_{k,2}-\beta x_{k,3}\big)$, $h=0.01$, with output $y_k=x_{k,1}$,
$\theta=(\sigma,\rho,\beta)$, $\theta^\star=(10,28,8/3)$ and
$x_0^\star=(-8,7,27)$; it is chaotic at $\theta^\star$.

A random start with spreads $\Delta_\theta,\Delta_x>0$ draws
$\theta^0=\theta^\star+\Delta_\theta\xi_\theta$ and
$x_0^0=x_0^\star+\Delta_x\xi_x$, with $\xi_\theta,\xi_x$ independent standard
Gaussian vectors, and sets $z$ to the free run of the model from
$(\theta^0,x_0^0)$ and $d=0$, so that no knowledge of the true states is
used. Starts whose free-run output exceeds ten times the largest $|y_k|$ of the
record are redrawn, for all methods. Gradient descent starts from the step $\eta=1/\hat\ell$, with $\hat\ell$ an
estimate of the Lipschitz constant of the gradient near the start, and halves
$\eta$ whenever $L(w^+)\le L(w)-\tfrac{\eta}{2}\norm{\nabla L(w)}^2$ fails. 
Algorithm~\ref{alg:homotopy} runs $1200$ gradient steps on $L_0$ and
then $T=8$ reductions from $\gamma_0=1$ to $\gamma_\star$, with
$n_j=150$ iterations per stage. Each stage takes Levenberg--Marquardt steps~\cite{levenberg1944LeastSquares,marquardt1963LeastSquares} on the analytic $\Jr_\gamma$,
stopped when $\norm{\Jr_\gamma\trans r_\gamma}<10^{-12}$, instead of the
gradient steps of Proposition~\ref{prop:homotopy}, so the experiments test the
continuation mechanism rather than its rate. The baselines are \emph{single shooting} (SS), $1350$ Levenberg--Marquardt
iterations on $L_{\rm sim}$, the budget of Algorithm~\ref{alg:homotopy}, and
three solves at the target weight without continuation: the direct solve (D),
$300$ Levenberg--Marquardt iterations on $L_{\gamma_\star}$ from the random
start; the fit phase followed by D (F+D); and the fit phase followed by D with
the budget of Algorithm~\ref{alg:homotopy} (F+D$_{1350}$). A run recovers
$\theta^\star$ if its estimate $\hat\theta$ satisfies
$\norm{\hat\theta-\theta^\star}<10^{-4}$. Records are noiseless unless stated
otherwise.

\subsection{Horizon-uniform geometry of the fit loss}
\label{subsec:num-geometry}

By Theorem~\ref{thm:pl} and Corollary~\ref{cor:cond}, the floor
$\lmin(K)\ge\mu/2$ should not depend on $N$, and $\kappa(K)$ should grow at
most linearly in $N$, through the parameter term only, whereas the curvature
of single shooting grows geometrically when the dynamics expand, by
\eqref{eq:cond}. Table~\ref{tab:cond} evaluates both at the true
configuration \eqref{eq:wstar} of the synthetic benchmark. The floor holds in
every configuration and is tight within a factor two,
$\lmin(K)/(\mu/2)\in[1.70,1.90]$, uniformly in $N$. The condition number
$\kappa(K)$ grows linearly in $N$ when the parameter term dominates
($b_0=2$) and is flat otherwise. For single shooting with $a_0=1.53$,
$\kappa(\partial\Ups)^2$ reaches $10^{16}$
already at $N=40$ and saturates beyond, so we measure the growth on $\norm{\partial\Ups}^2$,
the quantity bounded in \eqref{eq:cond}: a least-squares fit of
$\log_{10}\norm{\partial\Ups}^2$ against $N\in\{10,\dots,160\}$ has slope
$0.370$, against $2\log_{10}a_0=0.369$ predicted by \eqref{eq:cond} with
$\underline a\approx a_0$. The floor also depends on the output bound: by Remark~\ref{rem:reading},
$\mu=\underline c^{\,2}/(\Omega+1)$ on the second branch of
\eqref{eq:pl-const}. Over $14$ log-spaced values of $c_0=\underline c$ in
$[0.1,5]$ at $N=20$, for $a_0=0.53$ and $1.53$, the floor holds at every
point, with $\lmin(K)/(\mu/2)\ge1.04$, and on the second branch
$\lmin(K)\propto\underline c^{\,1.94}$ and $\underline c^{\,1.95}$, respectively.

\begin{table}[t]
\centering
\caption{Conditioning at $w^\star$ on the synthetic benchmark: range of
$\lmin(K)/(\mu/2)$ over $N\in\{10,\dots,160\}$, $\kappa(K)$ at the extreme
horizons and its fitted exponent in $N$ (exp.), and $\kappa(\partial\Ups)^2$
of single shooting at $N=10$ and $N=40$.}
\label{tab:cond}
\footnotesize
\setlength{\tabcolsep}{3.2pt}
\begin{tabular}{@{}ccccccccc@{}}
\toprule
 & & & \multicolumn{3}{c}{$\kappa(K)$} & & \multicolumn{2}{c}{$\kappa(\partial\Ups)^2$}\\
\cmidrule(lr){4-6}\cmidrule(l){8-9}
$b_0$ & $a_0$ & $\frac{\lmin(K)}{\mu/2}$ & $N{=}10$ & $N{=}160$ & exp. & & $N{=}10$ & $N{=}40$\\
\midrule
0.05 & 0.53 & 1.70--1.73 & 16.0 & 16.4 & 0.01 & & $9.3{\times}10^{4}$ & $3.7{\times}10^{4}$\\
0.05 & 0.93 & 1.78--1.81 & 29.0 & 29.8 & 0.01 & & $1.1{\times}10^{5}$ & $1.6{\times}10^{2}$\\
0.05 & 1.53 & 1.86--1.89 & 64.5 & 66.3 & 0.01 & & $2.4{\times}10^{6}$ & $1.2{\times}10^{16}$\\
\midrule
2.00 & 0.53 & 1.70--1.72 & 165 & 2595 & 0.99 & & $9.1{\times}10^{6}$ & $3.8{\times}10^{7}$\\
2.00 & 0.93 & 1.78--1.81 & 221 & 3501 & 1.00 & & $3.5{\times}10^{5}$ & $5.1{\times}10^{4}$\\
2.00 & 1.53 & 1.86--1.90 & 333 & 5222 & 0.99 & & $1.0{\times}10^{7}$ & $1.1{\times}10^{16}$\\
\bottomrule
\end{tabular}
\end{table}

\begin{figure}[t]
\centering
\includegraphics[width=\columnwidth]{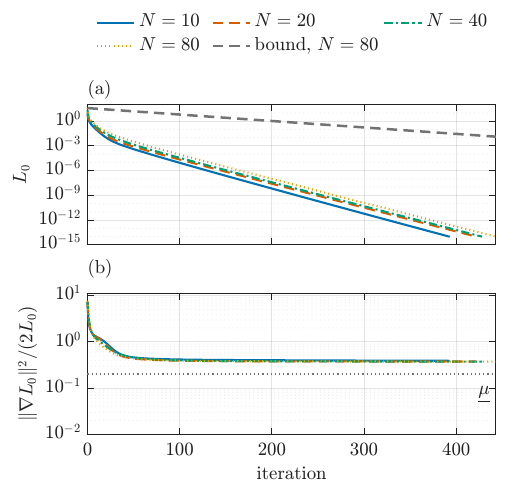}
\caption{Gradient descent on $L_0$, synthetic benchmark ($a_0=0.93$). (a)
$L_0(w^t)$ and, for $N=80$, the bound $(1-\underline\mu\eta)^tL_0(w^0)$ of
\eqref{eq:linconv}. (b) $\tfrac12\norm{\nabla L_0}^2/L_0$ and $\underline\mu$ (dotted).}
\label{fig:rate}
\end{figure}

Figure~\ref{fig:rate} tests Corollary~\ref{cor:linconv} away from $\Mset$.
Gradient descent on $L_0$ starts at $(\theta^\star+0.5\,\xi_\theta,\
x_0^\star+0.3\,\xi_x,\ 0.3\,\xi_z,\ 0)$, with $\xi_z$ a standard Gaussian
vector of the size of $z$, for $N\in\{10,20,40,80\}$ and a step common to all
horizons. The PL inequality holds at every iterate,
$\tfrac12\norm{\nabla L_0}^2/L_0\ge1.82\,\underline\mu$, with $\underline\mu$
the smallest value of $\mu(w^t)$ along each trajectory, as in
Corollary~\ref{cor:linconv}, and the linear rates differ by at most $6\%$
across horizons while exceeding the rate $\underline\mu\eta=0.018$ of
\eqref{eq:linconv} by about a factor four.

\subsection{Comparison with single shooting}
\label{subsec:num-horizon}

The horizon-uniform floor has a practical consequence. Table~\ref{tab:horizon}
compares single shooting and Algorithm~\ref{alg:homotopy} with
$\gamma_\star=10^{-6}$ from the same $40$ random starts, with spreads $\Delta_\theta=1$ and
$\Delta_x=0.5$ ($0.3$ and $0.15$ for the lightly damped oscillator), on records of
increasing length. At the truth, $\lmin(K)$ varies by less than $2\%$ with
the horizon, whereas $\kappa(\partial\Ups)^2$ reaches $10^{16}$ on the expanding
synthetic benchmark and $10^{10}$ on the Lorenz benchmark. Single shooting
recovers $\theta^\star$ on short records but fails increasingly often as $N$
grows, down to no run at $N=40$ and $N=800$ on these two benchmarks, while
Algorithm~\ref{alg:homotopy} recovers it in $39$ or $40$ runs at every
horizon. On the lightly damped oscillator, whose conditioning stays moderate,
single shooting loses only $5$ runs at $N=800$. The fit phase followed by
a single solve at $\gamma_0=1$ recovers $\theta^\star$ in the same runs as Algorithm~\ref{alg:homotopy}, as
expected, since $w^\star$ minimizes $L_\gamma$ at every weight (Theorem~\ref{thm:regpath}(i)).
\begin{table}[t]
\centering
\caption{Recoveries of $\theta^\star$ in $40$ random starts on noiseless
records: single shooting (SS) and Algorithm~\ref{alg:homotopy} (A1,
$\gamma_\star=10^{-6}$); $\kappa(\partial\Ups)^2$ and $\lmin(K)$ at the
truth.}
\label{tab:horizon}
\footnotesize
\setlength{\tabcolsep}{4.5pt}
\begin{tabular}{@{}lccccc@{}}
\toprule
benchmark & $N$ & $\kappa(\partial\Ups)^2$ & $\lmin(K)$ & SS & A1\\
\midrule
synthetic, $a_0=1.53$ & 10 & $2.4{\times}10^{6}$ & 0.125 & 40 & 39\\
 & 20 & $2.8{\times}10^{9}$ & 0.123 & 40 & 40\\
 & 30 & $3.0{\times}10^{12}$ & 0.123 & 27 & 40\\
 & 40 & $1.2{\times}10^{16}$ & 0.123 & \phantom{0}0 & 40\\
\midrule
Duffing, $\delta^\star=0.05$ & 100 & $1.6{\times}10^{4}$ & 0.172 & 40 & 40\\
 & 200 & $2.7{\times}10^{4}$ & 0.172 & 40 & 40\\
 & 400 & $2.8{\times}10^{4}$ & 0.172 & 37 & 40\\
 & 800 & $5.2{\times}10^{4}$ & 0.172 & 35 & 40\\
\midrule
Lorenz & 100 & $1.1{\times}10^{5}$ & 0.184 & 40 & 40\\
 & 200 & $3.3{\times}10^{4}$ & 0.184 & 33 & 40\\
 & 400 & $1.4{\times}10^{6}$ & 0.184 & 15 & 39\\
 & 800 & $9.4{\times}10^{9}$ & 0.184 & \phantom{0}0 & 40\\
\bottomrule
\end{tabular}
\end{table}

\begin{figure*}[!t]
\centering
\includegraphics[width=\textwidth]{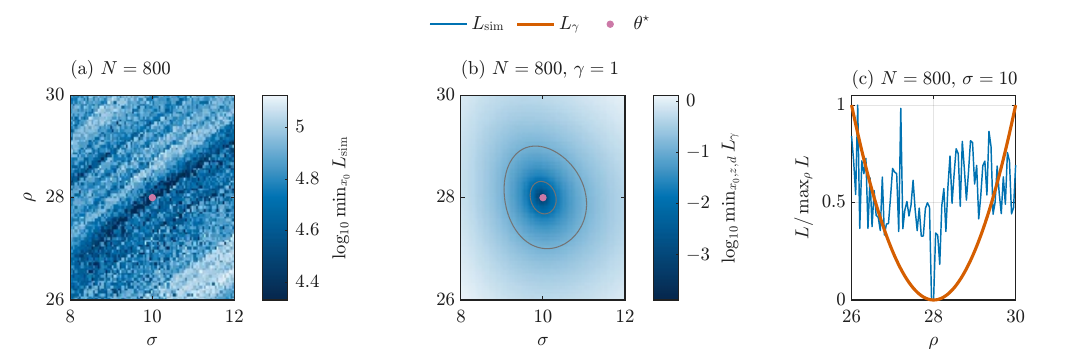}
\caption{Lorenz benchmark at $N=800$ and $\beta=8/3$: (a) $\min_{x_0}L_{\rm sim}$;
(b) $\min_{x_0,z,d}L_\gamma$ at $\gamma=1$, with level sets at $10^{-1}$ and
$10^{-2}$ times its maximum; (c) the two minimized losses $L$ of (a) and (b)
along $\sigma=10$, each divided by its maximum over $\rho$.}
\label{fig:landscape}
\end{figure*}

Figure~\ref{fig:landscape} shows both losses on the Lorenz benchmark at $N=800$,
on the plane $(\sigma,\rho)$ at $\beta=8/3$, within two standard
deviations of the random starts, each minimized over its remaining variables on
the same $81\times81$ grid. The map of $L_{\rm sim}$ has $687$ local minima, and
$176$ on the $41\times41$ subgrid;
$\min_{x_0,z,d}L_\gamma$ at $\gamma=1$ has one at both resolutions, at
$\theta^\star$. 

\subsection{Zero loss, selection, and continuation}
\label{subsec:num-selection}

\begin{figure}[!t]
\centering
\includegraphics[width=\columnwidth]{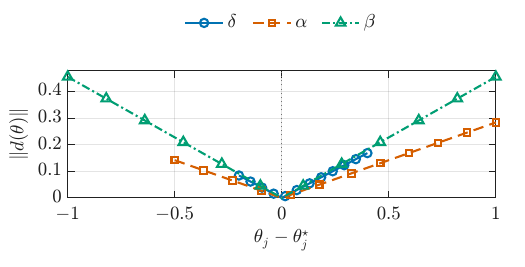}
\caption{Zero-loss completions on the Duffing benchmark ($N=40$): correction
norm $\norm{d(\theta)}$ along sweeps of $\delta$, $\alpha$ and $\beta$
through $\theta^\star$; $L_0$ is zero to machine precision at every point.}
\label{fig:curse}
\end{figure}

Figure~\ref{fig:curse} illustrates Proposition~\ref{prop:curse} on the
Duffing benchmark at $N=40$. Along sweeps of each coordinate of $\theta$
through $\theta^\star$, the completion $(\theta,x_0^\star,z^\star,d(\theta))$
of the proof of Proposition~\ref{prop:curse} has $L_0=0$, whereas its correction norm is smallest at the grid point nearest
to $\theta^\star$ and grows linearly away from it. The loss thus carries no information about $\theta$ on these completions, whereas the correction norm does, and Theorem~\ref{thm:exact} shows that its minimizer over all of $\Mset$ is the truth.

The penalized loss $L_\gamma$ performs this selection, but with little
curvature. By Proposition~\ref{prop:localPL}(ii), its curvature at $w^\star$
is $\gamma\rho_\star$ for small $\gamma$; at $N=20$, on the Duffing and
reparameterized benchmarks, $\lmin(\Jr_\gamma\trans\Jr_\gamma)/(\gamma\rho_\star)$
equals $1$ to five significant digits at $\gamma=10^{-6}$, with
$\rho_\star=1.4\times10^{-6}$ and $1.6\times10^{-6}$. On a short record, weak
identifiability thus leaves little curvature for selection, and the curvature decreases
in proportion to the weight.

Table~\ref{tab:ablation} shows the consequence for a solve at the target
weight and tests continuation, at $N=20$, from $40$ random starts with
$\Delta_\theta=1$ and $\Delta_x=0.5$.
Algorithm~\ref{alg:homotopy} recovers $\theta^\star$ in every run, at every
target weight and on both benchmarks, with at most $283$ Levenberg--Marquardt iterations over the nine stages. The direct solve fails increasingly often as $\gamma_\star$
decreases, on the reparameterized benchmark already at $10^{-2}$, and neither
the fit phase nor a larger budget closes the gap: F+D$_{1350}$ fails at
$\gamma_\star=10^{-8}$ in $39$ and $35$ runs. Every failed run ends at its
budget numerically on $\Mset$, with $L_0<4\times10^{-9}$ and a wrong
parameter. On $\Mset$ the gradient of $L_{\gamma_\star}$ reduces to
its $d$-block $2\gamma_\star d$, and at $\gamma_\star=10^{-8}$ the curvature at
$w^\star$ is $\gamma_\star\rho_\star\approx10^{-14}$, so a solve at the
target weight barely moves along $\Mset$; continuation hands each weight a start
produced at a larger one. 
\begin{table}[t]
\centering
\caption{Recoveries of $\theta^\star$ in $40$ random starts ($N=20$): direct
solve (D), fit phase followed by D (F+D), Algorithm~\ref{alg:homotopy}
(A1), and fit phase followed by D with the budget of A1
(F+D$_{1350}$).}
\label{tab:ablation}
\footnotesize
\setlength{\tabcolsep}{4.5pt}
\begin{tabular}{@{}lccccc@{}}
\toprule
system & $\gamma_\star$ & D & F+D & A1 & F+D$_{1350}$\\
\midrule
Duffing & $10^{-2}$ & 40 & 40 & 40 & 40\\
Duffing & $10^{-4}$ & 35 & 35 & 40 & 40\\
Duffing & $10^{-6}$ & \phantom{0}2 & \phantom{0}2 & 40 & 36\\
Duffing & $10^{-8}$ & \phantom{0}0 & \phantom{0}0 & 40 & \phantom{0}1\\
\midrule
reparam. & $10^{-2}$ & 23 & 24 & 40 & 40\\
reparam. & $10^{-4}$ & 19 & 19 & 40 & 24\\
reparam. & $10^{-6}$ & 11 & \phantom{0}9 & 40 & 17\\
reparam. & $10^{-8}$ & \phantom{0}1 & \phantom{0}2 & 40 & \phantom{0}5\\
\bottomrule
\end{tabular}
\end{table}

The fold benchmark shows that the hypothesis of
Proposition~\ref{prop:spurious} can hold. Since $\theta$ enters $f$ only
through $\alpha$, and affinely in it, the completions of
Proposition~\ref{prop:curse} have $d(\theta)=(\alpha(\theta)-1)\,u$ for a
fixed vector $u$, so $\norm{d(\theta)}^2=(\alpha(\theta)-1)^2\norm{u}^2$ has a
local maximum at $\theta=1$ and a nonglobal local minimum at $\theta=-1$. At
$\gamma=10^{-3}$ and $N=20$, $L_\gamma$ has a strict local minimizer at
$\theta=-1.0000$, with $\norm{d}=6.1\times10^{-2}$. From $60$ random starts with
spreads $\Delta_\theta=2$ and $\Delta_x=0.2$, with $2000$ iterations for the
direct solve and $1500$ fit steps and $500$ iterations per stage for
Algorithm~\ref{alg:homotopy}, the $13$ with $\theta^0<1$ end
at $\theta=-1$, both for the direct solve and for
Algorithm~\ref{alg:homotopy}. Of the other $47$, the direct solve recovers
$\theta^\star$ in $33$ and stalls near $\theta=1$ in $14$, whereas
Algorithm~\ref{alg:homotopy} recovers $\theta^\star$ in all of them.
Continuation recovers every start on the side of $\theta^\star$, including
those on which the direct solve stalls,  while the $13$ starts with $\theta^0<1$ still end at $\theta=-1$, as Remark~\ref{rem:homotopyscope} anticipates.

\subsection{Robustness to mismatch}
\label{subsec:num-robust}

\begin{figure}[t]
\centering
\includegraphics[width=\columnwidth]{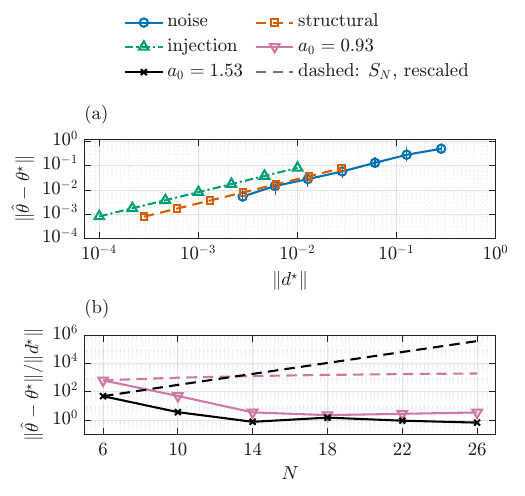}
\caption{Robust recovery, $\hat\theta$ from Algorithm~\ref{alg:homotopy}
($\gamma_\star=10^{-6}$). (a) Duffing benchmark, $N=40$: error versus
$\norm{d^\star}$ (output noise: median and interquartile range over $40$
realizations). (b) Synthetic benchmark with an injected correction of norm $10^{-5}$:
normalized error $\norm{\hat\theta-\theta^\star}/\norm{d^\star}$ versus $N$
(solid) for $a_0=0.93$ and $1.53$, and the horizon factor $S_N$ of
\eqref{eq:SN}, multiplied by a constant that makes it equal to the normalized error at $N=6$ (dashed).}
\label{fig:robust}
\end{figure}

Figure~\ref{fig:robust}(a) tests Theorem~\ref{thm:robust} on the Duffing
benchmark at $N=40$, for three mechanisms with a true configuration
\eqref{eq:truecorrected} built explicitly, so that $\norm{d^\star}$ is known:
output noise $y_k=q_k^\star+\nu_k$, $\nu_k\sim\mathcal N(0,\varsigma^2)$,
absorbed by the measured coordinate of the lifted state; an unmodeled
stiffness $-\beta_5q^5$ in the data-generating system; and a correction
$d^\star$ injected along a fixed unit direction. The levels $\varsigma$,
$\beta_5\in[10^{-3.5},10^{-1.5}]$ and $\norm{d^\star}\in[10^{-4},10^{-2}]$
take seven logarithmically spaced values, with $40$ noise realizations per
level. The estimate $\hat\theta$ is returned by
Algorithm~\ref{alg:homotopy} with $\gamma_\star=10^{-6}$ and $200$ iterations
per stage, from random starts with spreads $\Delta_\theta=0.3$ and
$\Delta_x=0.2$; on the exact record its error is
$1.1\times10^{-12}$. The error is linear in $\norm{d^\star}$, with fitted
exponents $0.98$, $1.00$ and $1.00$. 

Figure~\ref{fig:robust}(b) tests whether the error follows the horizon factor
$S_N$ of Theorem~\ref{thm:robust}, which is bounded uniformly in $N$ under
contraction and grows with $N$ otherwise. On the synthetic benchmark, a
correction of norm $10^{-5}$ is injected, with $N\in\{6,10,\dots,26\}$ and
$a_0=0.93$ and $1.53$, i.e., $a(w^\star)=0.96$ and $1.56$; the starts perturb
$\theta^\star$, $x_0^\star$ and $z^\star$ by Gaussian vectors with standard
deviations $0.2$, $0.1$ and $0.05$, with $1500$ fit steps and $200$ iterations
per stage. The normalized error stays below its value at $N=6$ for both $a(w^\star)=0.96$
and $a(w^\star)=1.56$, whereas $S_N$ grows by almost four orders of magnitude in the
expanding case. On this benchmark the error is thus horizon-uniform in both
regimes, and $S_N$ is conservative for expanding dynamics.

On records that are not realizable, the minimizer of $L_{\gamma_\star}$
depends on $\gamma_\star$; we thus finally test whether Algorithm~\ref{alg:homotopy}
reaches it and how accurate it is, under process noise: on the Duffing benchmark at $N=200$, a zero-mean Gaussian
perturbation with covariance $\varsigma_{\rm p}^2I$ is added to each state
update ($\varsigma_{\rm p}=10^{-3}$, $10^{-2}$), with $40$ realizations per
level and one random start each ($\Delta_\theta=1$, $\Delta_x=0.5$). The
minimizer of $L_{\gamma_\star}$ near the truth is no longer $w^\star$ and has
no closed form; as reference, we compute it by Levenberg--Marquardt iterations
started at $w^\star$ of \eqref{eq:truecorrected}, which is known on simulated
records. At $\gamma_\star=10^{-8}$, Algorithm~\ref{alg:homotopy} reaches the
reference in $39$ and $40$ runs, with median errors $7.2\times10^{-3}$ and
$1.0\times10^{-1}$, against $1.1\times10^{-2}$ and $1.6\times10^{-1}$ for single
shooting from the same starts, which reaches its own minimizer in $35$ runs at both levels.

\section{Conclusions}
\label{sec:conclusions}

This paper has addressed simulation-error identification of nonlinear
state-space models on long records, where the loss is nonconvex and its
conditioning can degrade geometrically with the horizon. We lifted the
trajectory with a free, softly penalized correction at each step and showed
that the fit loss of the lifted problem satisfies a PL inequality whose
constant does not depend on the horizon (Theorem~\ref{thm:pl}). The price
is identifiability, since a perfect fit no longer determines the physical
parameter (Proposition~\ref{prop:curse}); we restored it by selecting the
perfect fit with minimal correction, which the global minimizer of the
penalized loss returns at every positive weight and which recovers the true
parameters exactly under structural identifiability, with an error linear
in the unmodeled effects otherwise
(Theorems~\ref{thm:exact}--\ref{thm:robust}). The resulting estimator,
gradient descent on the fit loss followed by a homotopy in the penalty
weight, combines the two steps with a local convergence guarantee
(Proposition~\ref{prop:homotopy}).

In the experiments, the lifted estimator recovered the parameters on long
records of expanding and chaotic systems on which single shooting failed. A
single solve at a moderate weight after the fit phase did as well as
continuation, which is needed to reach a small target weight, where a single solve stalls.

Two limits are intrinsic: the penalized loss can have spurious local
minimizers (Proposition~\ref{prop:spurious}), and continuation still needs a
start in the basin of the initial weight (Remark~\ref{rem:homotopyscope}).
Open directions are an extension of the selection results to kernel-norm
penalties on the correction, which would carry these guarantees to
kernel-based physics-informed identification~\cite{donati2026kernel},
an analysis of continuation without exact realizability, conditions guaranteeing a priori that the iterates remain where the
assumptions hold, and a robustness bound that is horizon-uniform without
incremental contraction, as the experiments suggest.

\appendices
\section{Notation}
\label{app:notation}
Table~\ref{tab:notation} collects the main symbols of the paper, grouped by topic, with the place where each is defined.

\begin{table}[!htbp]
\centering
\caption{Notation table.}
\label{tab:notation}
\small
\setlength{\tabcolsep}{3pt}
\renewcommand{\arraystretch}{1.05}
\begin{tabular}{@{}l>{\raggedright\arraybackslash}p{9.5cm}l@{}}
\toprule
Symbol & Meaning & Defined in\\
\midrule
\multicolumn{3}{@{}l}{\emph{Model and data}}\\
$x_k$, $y_k$, $\theta$ & state, output, parameter; dimensions $\nx$, $\ny$, $\nth$ & \eqref{eq:system}\\
$N$, $y$ & horizon; record $(y_1,\dots,y_N)$ & \eqref{eq:simerror}\\
$\fk{k}$ & $k$-fold iterate of $f(\cdot,\theta)$ & Sec.~\ref{sec:simerror}\\
$\Ups$, $\Ups_k$ & simulated output trajectory, its blocks & \eqref{eq:Upsilon}\\
$L_{\mathrm{sim}}$ & simulation error & \eqref{eq:simerror}\\
\multicolumn{3}{@{}l}{\emph{Lifted problem}}\\
$z$, $d$ & lifted state ($z_0\eqdef x_0$), per-step correction & Sec.~\ref{subsec:lifted}\\
$w$, $p$ & decision vector $(\theta,x_0,z,d)\in\R^p$, $p=\nth+\nx+2N\nx$ & \eqref{eq:param-vector}\\
$e$, $s$ & consistency and output residuals & \eqref{eq:res-e}, \eqref{eq:res-s}\\
$r$, $m$ & stacked residual $(e,s)\in\R^m$, $m=N(\nx+\ny)$ & \eqref{eq:losses}\\
$L_0$, $L_\gamma$ & fit loss; penalized loss with weight $\gamma\ge0$ & \eqref{eq:losses}\\
$p-m$ & overparameterization margin & \eqref{eq:margin}\\
\multicolumn{3}{@{}l}{\emph{Optimization geometry}}\\
$J$, $K$ & residual Jacobian $\partial r$, tangent kernel $JJ\trans$ & Def.~\ref{def:kernel}, \eqref{eq:Jacobian}\\
$A_k$, $B_k$, $C_k$ & one-step sensitivities along the lifted states & \eqref{eq:sens}\\
$a(w)$, $c(w)$ & $\max_k\norm{A_k}$ and $\min_k\smin(C_k)$ at $w$ & \eqref{eq:scalars}\\
$\Wset$ & region of interest in $\R^p$ & Sec.~\ref{subsec:pl-theorem}\\
$\underline c$, $\bar a$ & uniform bounds $c(w)\ge\underline c$, $a(w)\le\bar a$ on $\Wset$ & Ass.~\ref{ass:submersion}, \ref{ass:bounded}\\
$\mu$, $\Omega$ & horizon-uniform PL constant; $\Omega=(1+\bar a)^2$ & \eqref{eq:pl-const}\\
$\mu(w)$ & $\mu$ evaluated at $(c(w),a(w))$ & Sec.~\ref{subsec:pl-theorem}\\ $\underline\mu$ & floor of $\mu(w)$ along the iterates & Cor.~\ref{cor:linconv}\\
$\ell_0$, $\eta$ & Lipschitz modulus of $\nabla L_0$; step size of gradient descent & Cor.~\ref{cor:linconv}\\
$\Mset$ & zero-loss set $\{w\in\Wset:L_0(w)=0\}$ & \eqref{eq:zeroloss_set}\\
\multicolumn{3}{@{}l}{\emph{Identifiability and selection}}\\
$w^\star$ & true configuration $(\theta^\star,x_0^\star,z^\star,d^\star)$; $d^\star=0$ under Ass.~\ref{ass:realiz} & \eqref{eq:wstar}, \eqref{eq:truecorrected}\\
$\pi_\bullet$ & projection of a set onto the named blocks & Sec.~\ref{sec:curse}\\
$\chi$ & physical pair $(\theta,x_0)$ & Sec.~\ref{sec:curse}\\
$w^\dagger$ & minimal-correction selector & \eqref{eq:selector}\\
$w_\gamma$ & minimizer of $L_\gamma$ over $\Wset$ & Thm.~\ref{thm:regpath}\\
$\rho_\star$, $\sigma_\star$ & $\min\norm{v_d}^2$ over unit $v\in\ker J(w^\star)$; $\smin(\partial\Ups(\chi^\star))$ & Prop.~\ref{prop:localPL}\\
$\gamma_{\max}$ & largest admissible weight, $\gamma_0\le\gamma_{\max}<\mu/2$ & Lemma~\ref{lem:scLgamma}\\
$\sigma_0$, $\varrho$, $\mathcal B_\gamma$ & strong-convexity factor, radius factor, ball $\{\norm{w-w^\star}\le\varrho\gamma\}$ & Lemma~\ref{lem:scLgamma}\\ $\ell_H$, $\ell_B$ & Hessian modulus around $w^\star$; smoothness modulus of $L_\gamma$ on $\mathcal B_\gamma$ & Lemma~\ref{lem:scLgamma}\\
$\gamma_0$, $\gamma_\star$, $T$, $\tau$, $n_j$ & initial and target weights, number of reductions, ratio of the schedule $\gamma_j=\gamma_0\tau^j$, steps per stage & Sec.~\ref{subsec:homotopy}\\
$\underline\upsilon$ & lower modulus of $\Ups$ at $\chi^\star$ & \eqref{eq:lowermodulus}\\
$\bar a_+$, $\ell_g$, $S_N$ & sensitivity bounds of the robustness analysis & \eqref{eq:SN}\\
\bottomrule
\end{tabular}
\end{table}

\section{Proofs}
\label{app:proofs}
\subsection{Proof of Theorem~\ref{thm:pl}}\label{app:pf-thm-pl}
\begin{proof}
We first bound $D_f$, at any $w$. Write $D_f=\bI-\Sigma$, where $\Sigma$ is the strictly block-lower-bidiagonal matrix whose
$(k,k-1)$ block equals $A_k$ for $k=2,\dots,N$ and which is zero otherwise. The nonzero
blocks of $\Sigma$ occupy distinct block rows and distinct block columns, hence
$\Sigma\,\Sigma\trans=\blkdiag\big(\bzero,\,A_2A_2\trans,\,\dots,\,A_NA_N\trans\big)$, whose largest
eigenvalue is $\max_{2\le k\le N}\norm{A_k}^2\le a(w)^2$. Therefore $\norm \Sigma\le a(w)$ and,
by the triangle inequality,
\begin{equation}
  \norm{D_f}\le\norm{\bI}+\norm \Sigma\le 1+a(w),
  \label{eq:Df}
\end{equation}
which does not involve~$N$.
By Lemma~\ref{lem:pl}, it now suffices to show $\lmin(K(w))\ge\mu/2$. Fix
$w\in\Wset$ and a unit vector $u=(u_e,u_s)\in\R^{N\nx}\times\R^{N\ny}$. The
correction columns of \eqref{eq:Jacobian} contribute $-u_e$ to $\Jr\trans u$
and the $z$-columns contribute $D_f\trans u_e-G\trans u_s$; discarding the
nonnegative contribution of the $(\theta,x_0)$-columns,
\begin{equation}
  u\trans K\,u=\norm{\Jr\trans u}^2
  \ge \norm{u_e}^2 + \norm{D_f\trans u_e - G\trans u_s}^2 .
  \label{eq:proof-lb}
\end{equation}
For $\varepsilon\in(0,1)$, the Young inequality
$2\inner{D_f\trans u_e}{G\trans u_s}\le\tfrac1\varepsilon\norm{D_f\trans u_e}^2+\varepsilon\norm{G\trans u_s}^2$
gives
\begin{equation*}
  \norm{D_f\trans u_e-G\trans u_s}^2\ge\big(1-\tfrac1\varepsilon\big)\norm{D_f\trans u_e}^2+(1-\varepsilon)\norm{G\trans u_s}^2 .
\end{equation*}
We use
$\norm{D_f\trans u_e}^2\le(1+a(w))^2\norm{u_e}^2\le\Omega\norm{u_e}^2$
by \eqref{eq:Df} and Assumption~\ref{ass:bounded}, together with
$\norm{G\trans u_s}^2\ge\underline c^{\,2}\norm{u_s}^2$, which holds because
$G$ is block diagonal with $\smin(G)=c(w)\ge\underline c$
(Assumption~\ref{ass:submersion}). Since $1-\tfrac1\varepsilon<0$,
\begin{equation}
  u\trans K\,u \ge
  \big[1-\big(\tfrac1\varepsilon-1\big)\Omega\big]\norm{u_e}^2
  + (1-\varepsilon)\,\underline c^{\,2}\,\norm{u_s}^2 .
  \label{eq:proof-coeffs}
\end{equation}
The choice $\varepsilon=\frac{2\Omega+1}{2\Omega+2}$ gives
$\tfrac1\varepsilon-1=\frac{1}{2\Omega+1}$, so the two coefficients equal
$\frac{\Omega+1}{2\Omega+1}$ and $\frac{\underline c^{\,2}}{2(\Omega+1)}$,
both not smaller than $\mu/2$. Hence
$u\trans Ku\ge\tfrac\mu2(\norm{u_e}^2+\norm{u_s}^2)=\tfrac\mu2$, and
$\lmin(K)\ge\mu/2$ because $u$ is arbitrary.
\end{proof}

\subsection{Proof of Corollary~\ref{cor:cond}}
\label{app:pf-cor-cond}
\begin{proof}
The lower bound on $\lmin(K)$ is Theorem~\ref{thm:pl}; the upper bound is the
Rayleigh quotient of $K$ at a unit vector $u=(0,u_s)$ supported on the output
rows, for which $\norm{\Jr\trans u}=\norm{G\trans u_s}\le\bar c$. For any
partition of $\Jr$ into column blocks $M_i$, the identity
$\norm{\Jr\trans u}^2=\sum_i\norm{M_i\trans u}^2$ gives
$\lmax(K)=\norm{\Jr}^2\le\sum_i\norm{M_i}^2$. The four column blocks of
\eqref{eq:Jacobian} have squared norms $\norm{B_e}^2$,
$\norm{X_e}^2=\norm{A_1}^2\le\bar a^2$, at most
$\norm{D_f}^2+\norm{G}^2\le(1+\bar a)^2+\bar c^{\,2}$ (bound \eqref{eq:Df}),
and $1$. Finally, $\lmax(K)\ge\norm{B_e}^2$, and $\norm{B_e}^2=\lmax\big(\sum_kB_k\trans B_k\big)$ is at least
the average eigenvalue $\nth^{-1}\operatorname{tr}\big(\sum_kB_k\trans B_k\big)\ge\nth^{-1}\sum_k\norm{B_k}^2$,
and at most $\sum_k\norm{B_k}^2=\Lambda_\theta$, by subadditivity of $\lmax$.
\end{proof}

\subsection{Proof of Corollary~\ref{cor:linconv}}
\label{app:pf-cor-linconv}
\begin{proof}
The segment joining $w^t$ and $w^{t+1}$ lies in the convex hull of the
iterates, so the descent lemma and $\eta\le1/\ell_0$ give
$L_0(w^{t+1})\le L_0(w^t)-\tfrac{\eta}{2}\norm{\nabla L_0(w^t)}^2$, and
Lemma~\ref{lem:pl} with $\lmin(K(w^t))\ge\mu(w^t)/2\ge\underline\mu/2$ gives
$L_0(w^{t+1})\le(1-\underline\mu\eta)L_0(w^t)$. If $L_0(w^t)>0$, then $1-\underline\mu\eta\ge0$
because $L_0(w^{t+1})\ge0$; if $L_0(w^t)=0$, the iterates are stationary from
$t$ on. Iterating yields \eqref{eq:linconv}. The descent inequality also gives
$\norm{w^{t+1}-w^t}=\eta\norm{\nabla L_0(w^t)}\le\sqrt{2\eta L_0(w^t)}
\le\sqrt{2\eta L_0(w^0)}\,(1-\underline\mu\eta)^{t/2}$, so the steps are summable, the
iterates converge to some $w^\infty$, and $L_0(w^\infty)=0$ by continuity. If the
iterates lie in a closed $\Wset$ on which Assumptions~\ref{ass:submersion}
and~\ref{ass:bounded} hold, then $\mu(w^t)\ge\mu$ of \eqref{eq:pl-const} and
$w^\infty\in\Wset$, hence $w^\infty\in\Mset$.
\end{proof}

\subsection{Proof of Theorem~\ref{thm:regpath}(ii)}
\label{app:pf-thm-regpath}
\begin{proof}
For any $w'\in\Mset$, optimality of $w_\gamma$,
i.e., $L_0(w_\gamma)+\gamma\norm{d_\gamma}^2\le\gamma\norm{d'}^2$,
together with $L_0(w_\gamma)\ge0$, gives
\begin{equation}
  L_0(w_\gamma)\le\gamma\norm{d'}^2
  \quad\text{and}\quad
  \norm{d_\gamma}^2\le\norm{d'}^2 .
  \label{eq:regpath-two}
\end{equation}
Since $\Wset$ is compact, every sequence $\gamma_i\downarrow0$ has a
subsequence along which $w_{\gamma_i}$ converges in $\Wset$, so accumulation
points exist. Let $\tilde w=\lim_iw_{\gamma_i}$ be one of them. By continuity,
the first bound in \eqref{eq:regpath-two} gives $L_0(\tilde w)=0$, i.e.,
$\tilde w\in\Mset$, and the second gives $\norm{\tilde d}^2\le\norm{d'}^2$ for
every $w'\in\Mset$. Thus $\tilde w$ solves \eqref{eq:selector}.
\end{proof}

\subsection{Proof of Proposition~\ref{prop:inexactPL}}
\label{app:pf-prop-inexactPL}
\begin{proof}
Since
$\nabla L_\gamma=\nabla L_0+2\gamma\,(0,0,0,d)$ with
$\norm{2\gamma\,(0,0,0,d)}\le2\gamma M_d$ on $\Wset$, the bound $\norm{\nabla L_0}^2\le2\norm{\nabla L_\gamma}^2+8\gamma^2M_d^2$ and Theorem~\ref{thm:pl} give
$\tfrac12\norm{\nabla L_\gamma}^2\ge\tfrac14\norm{\nabla L_0}^2-2\gamma^2M_d^2
\ge\tfrac{\mu}{2}L_0-2\gamma^2M_d^2$. Substituting
$L_0=L_\gamma-\gamma\norm d^2\ge L_\gamma-\gamma M_d^2$ gives
\eqref{eq:inexactPL}, and both losses are nonnegative and vanish at
$w^\star$.
\end{proof}

\subsection{Proof of Corollary~\ref{cor:inexactPL}}
\label{app:pf-cor-inexactPL}
\begin{proof} Since $\ell_\Wset+2\gamma$ bounds the Lipschitz modulus of $\nabla L_\gamma$, the descent lemma on the segment joining $w^t$ and $w^{t+1}$ gives
$L_\gamma(w^{t+1})\le L_\gamma(w^t)-\tfrac{\eta}{2}\norm{\nabla L_\gamma(w^t)}^2$,
and \eqref{eq:inexactPL} yields
$L_\gamma(w^{t+1})\le(1-\tfrac{\eta\mu}{2})L_\gamma(w^t)+\eta\Xi_\gamma$, where
$1-\tfrac{\eta\mu}{2}\in[\tfrac23,1)$ because $\mu\le\tfrac43$ by
\eqref{eq:pl-const}, as $\Omega\ge1$, and $\eta\le1/(\ell_\Wset+2\gamma)\le\tfrac12$.
Unrolling and summing the geometric series gives the bound, and
$L_0\le L_\gamma$ gives
$\limsup_tL_0(w^t)\le2\Xi_\gamma/\mu=\gamma M_d^2+4\gamma^2M_d^2/\mu$.
\end{proof}

\subsection{Proof of Proposition~\ref{prop:spurious}}
\label{app:pf-prop-spurious}
\begin{proof}
Fix $\varepsilon>0$ so small that the closed ball $\bar B$ of radius
$\varepsilon$ centered at $\bar w$ lies in $\operatorname{int}\Wset$, does not
contain $w^\star$, and satisfies $\norm{d}^2>\norm{\bar d}^2$ on
$(\Mset\cap\bar B)\setminus\{\bar w\}$, and let $S$ be its boundary sphere. If
$\Mset\cap S\neq\emptyset$, it is compact, so $\norm{d}^2-\norm{\bar d}^2$ has a
positive minimum on it, and by uniform continuity there is $\varepsilon'>0$
such that $\norm{d}^2>\norm{\bar d}^2$ at every point of $S$ within distance
$\varepsilon'$ of $\Mset\cap S$. The remaining points of $S$ form a compact set
disjoint from $\Mset$, on which $L_0$ has a positive minimum $\underline L$.
Since $\bar d\neq0$ by Theorem~\ref{thm:exact}, for
$\gamma<\underline L/\norm{\bar d}^2$ every $w\in S$ satisfies either
$\gamma\norm{d}^2>\gamma\norm{\bar d}^2$ or
$L_0(w)\ge\underline L>\gamma\norm{\bar d}^2$, hence
$L_\gamma(w)>\gamma\norm{\bar d}^2=L_\gamma(\bar w)$. The minimum of $L_\gamma$
over $\bar B$ is therefore attained at an interior point $w'_\gamma$, a local
minimizer of $L_\gamma$. By Theorem~\ref{thm:exact}, $L_\gamma$ vanishes on
$\Wset$ only at $w^\star\notin\bar B$, so $L_\gamma(w'_\gamma)>0$, whereas a PL
inequality with reference value $\min_{\Wset}L_\gamma=0$ would force
$L_\gamma(w'_\gamma)=0$ at the stationary point $w'_\gamma$.
\end{proof}

\subsection{Proof of Proposition~\ref{prop:localPL}}
\label{app:pf-prop-localPL}
\begin{proof}
(i) A vector $v=(v_\theta,v_{x_0},v_z,v_d)$ lies in $\ker\Jr_\gamma(w^\star)$
if and only if $\Jr(w^\star)v=0$ and $v_d=0$. The consistency rows of
\eqref{eq:Jacobian} then read $B_ev_\theta+X_ev_{x_0}+D_fv_z=0$ and, since
$D_f$ is block lower-bidiagonal with identity diagonal, determine $v_z$
uniquely as the forward state sensitivity to $(v_\theta,v_{x_0})$; by the
chain rule, $Gv_z=\partial\Ups(\chi^\star)(v_\theta,v_{x_0})$, so the output
rows $-Gv_z=0$ read $\partial\Ups(\chi^\star)(v_\theta,v_{x_0})=0$. Thus $v\mapsto(v_\theta,v_{x_0})$ maps $\ker\Jr_\gamma(w^\star)$ injectively
into $\ker\partial\Ups(\chi^\star)$, which is $\{0\}$ by
Assumption~\ref{ass:ident}; hence $v_\theta=0$, $v_{x_0}=0$, then $v_z=0$, and
$v=0$.

(ii) With $\Gamma\eqdef\Jr(w^\star)\trans\Jr(w^\star)$ and the orthogonal
 projector $\Pi_d\eqdef\blkdiag(\bzero,\bzero,\bzero,\bI_{N\nx})$ onto the $d$-block, so that $\norm{\Pi_dv}=\norm{v_d}$, one has
$\Jr_\gamma(w^\star)\trans\Jr_\gamma(w^\star)=\Gamma+\gamma\Pi_d$. A unit
$v\in\ker\Gamma=\ker\Jr(w^\star)$ attaining $\rho_\star$ gives
$v\trans(\Gamma+\gamma\Pi_d)v=\gamma\rho_\star$, the upper bound. For the lower
bound, write a unit vector as $u=u_0+u_1$, with $u_0\in\ker\Gamma$ and
$u_1\perp\ker\Gamma$. The nonzero eigenvalues of $\Gamma$ are those of
$K(w^\star)$, hence at least $\mu/2$ by Theorem~\ref{thm:pl}, so
$u\trans\Gamma u\ge\tfrac\mu2\norm{u_1}^2$. Moreover,
$\norm{\Pi_du_0}^2\ge\rho_\star\norm{u_0}^2$, $\norm{\Pi_du_1}\le\norm{u_1}$, and the Young inequality
$2(\Pi_du_0)\trans \Pi_du_1\ge-\tfrac{2\gamma}{\mu}\norm{\Pi_du_0}^2-\tfrac{\mu}{2\gamma}\norm{\Pi_du_1}^2$
give, since $\gamma\le\mu/2$,
$u\trans\Pi_du=\norm{\Pi_du_0+\Pi_du_1}^2\ge(1-\tfrac{2\gamma}{\mu})\rho_\star\norm{u_0}^2
-(\tfrac{\mu}{2\gamma}-1)\norm{u_1}^2$. Hence
$u\trans(\Gamma+\gamma\Pi_d)u\ge\gamma\norm{u_1}^2
+\gamma\rho_\star(1-\tfrac{2\gamma}{\mu})\norm{u_0}^2
\ge\gamma\rho_\star(1-\tfrac{2\gamma}{\mu})$, since $\rho_\star\le1$. By (i),
no nonzero $v\in\ker\Jr(w^\star)$ satisfies $v_d=0$, so $\rho_\star>0$.

(iii) Let $(v_\theta,v_{x_0})$ be a unit right singular vector of
$\partial\Ups(\chi^\star)$ for $\sigma_\star$, let $v_z^0$ be its forward
state sensitivity, as in (i), so that $\norm{Gv_z^0}=\sigma_\star$, and let
$v_z\eqdef v_z^0-G^+Gv_z^0$, with $G^+\eqdef G\trans(GG\trans)^{-1}$, and
$v_d\eqdef B_ev_\theta+X_ev_{x_0}+D_fv_z=-D_fG^+Gv_z^0$. Then
$v\eqdef(v_\theta,v_{x_0},v_z,v_d)\in\ker\Jr(w^\star)$, $\norm v\ge1$, and
$\norm{v_d}\le\norm{D_f}\norm{G^+}\sigma_\star
\le(1+\bar a)\sigma_\star/\underline c$ by \eqref{eq:Df} and
Assumptions~\ref{ass:submersion}--\ref{ass:bounded}, so
$\rho_\star\le\norm{v_d}^2/\norm v^2\le\Omega\sigma_\star^2/\underline c^{\,2}$.
\end{proof}

\subsection{Proof of Lemma~\ref{lem:scLgamma}}
\label{app:pf-lem-scLgamma}
\begin{proof}
One has $\nabla^2L_0=2\Jr\trans\Jr+2\sum_i r_i\nabla^2r_i$, the sum running
over the components of $r$, and $\nabla^2L_0(w^\star)=2\Jr(w^\star)\trans\Jr(w^\star)$
because $r(w^\star)=0$. Let $\mathcal U$ be a closed ball centered at $w^\star$
on which $f$ and $g$ are twice continuously differentiable at the arguments
entering $r$. On $\mathcal U$, $\Jr$ is Lipschitz and the Hessians
$\nabla^2r_i$ are bounded, so
$\norm{\Jr(w)\trans\Jr(w)-\Jr(w^\star)\trans\Jr(w^\star)}$ and
$\norm{\sum_ir_i(w)\nabla^2r_i(w)}$ are both at most proportional to
$\norm{w-w^\star}$, which gives $\ell_H$. The penalty adds the constant matrix
$2\gamma\blkdiag(\bzero,\bzero,\bzero,\bI_{N\nx})$, so
$\nabla^2L_\gamma(w)\succeq\nabla^2L_\gamma(w^\star)-\ell_H\norm{w-w^\star}\bI$
on $\mathcal U$, and
$\nabla^2L_\gamma(w^\star)=2\Jr_\gamma(w^\star)\trans\Jr_\gamma(w^\star)\succeq2\sigma_0\gamma\bI$
for $\gamma\le\gamma_{\max}$ by Proposition~\ref{prop:localPL}(ii). Hence
$\nabla^2L_\gamma\succeq\sigma_0\gamma\bI$ on
$\mathcal B_\gamma\subseteq\mathcal B_{\gamma_{\max}}\subseteq\mathcal U$,
and $\norm{\nabla^2L_\gamma}\le\ell_B$ there. Strong convexity on the convex
set $\mathcal B_\gamma$ makes the minimizer unique, and it is $w^\star$
because $L_\gamma\ge0=L_\gamma(w^\star)$.
\end{proof}

\subsection{Proof of Proposition~\ref{prop:homotopy}}
\label{app:pf-prop-homotopy}
\begin{proof}
Let $w\in\mathcal B_{\gamma_j}$ and $w^+\eqdef w-\eta\nabla L_{\gamma_j}(w)$.
Since $\nabla L_{\gamma_j}(w^\star)=0$ and the segment from $w^\star$ to $w$
lies in the convex set $\mathcal B_{\gamma_j}$,
$w^+-w^\star=(\bI-\eta\bar H)(w-w^\star)$, where $\bar H$, the average of
$\nabla^2L_{\gamma_j}$ along that segment, has eigenvalues in
$[\sigma_0\gamma_j,\ell_B]$ by Lemma~\ref{lem:scLgamma}. As $\eta\le1/\ell_B$,
$\norm{w^+-w^\star}\le(1-\eta\sigma_0\gamma_j)\norm{w-w^\star}$, so
$w^+\in\mathcal B_{\gamma_j}$ and, after $n_j$ steps from
$w^{(j)}\in\mathcal B_{\gamma_j}$,
$\norm{w^{(j+1)}-w^\star}\le e^{-\eta\sigma_0\gamma_jn_j}\varrho\gamma_j
\le\tau\varrho\gamma_j=\varrho\gamma_{j+1}$ for $j<T$. Induction from
$w^{(0)}\in\mathcal B_{\gamma_0}$ gives the claims, the last because
$\theta^{(T+1)}$ is a block of $w^{(T+1)}$.
\end{proof}

\subsection{Proof of Lemma~\ref{lem:outsens}}
\label{app:pf-lem-outsens}
\begin{proof}
Let $\tilde z_k\eqdef\fk{k}(x_0,\theta)$, so that
$\tilde z_k=f(\tilde z_{k-1},\theta)$ and, by \eqref{eq:res-e},
$z_k=f(z_{k-1},\theta)+d_k+e_k$, both from $\tilde z_0=z_0=x_0$. The difference
$\zeta_k\eqdef z_k-\tilde z_k$ satisfies $\zeta_0=0$ and, by the mean-value
theorem in integral form, $\zeta_k=\bar A_k\zeta_{k-1}+d_k+e_k$ with
$\bar A_k\eqdef\int_0^1\ddx f(\tilde z_{k-1}+t\zeta_{k-1},\theta)\,dt$ and
$\norm{\bar A_k}\le\bar a_+$. Hence
$\norm{\zeta_k}\le\sum_{j=1}^{k}\bar a_+^{\,k-j}\norm{d_j+e_j}$, the discrete
convolution of $(\norm{d_j+e_j})_j$ with the kernel $(\bar a_+^{\,i})_{i\ge0}$,
and Young's convolution inequality gives
$\norm{\zeta}\le S_N\norm{d+e}\le S_N(\norm{d}+\norm{e})$. By \eqref{eq:res-s},
$\Ups_k(\chi)-y_k=g(\tilde z_k)-g(z_k)-s_k$, with
$\norm{g(\tilde z_k)-g(z_k)}\le\ell_g\norm{\zeta_k}$, and the triangle
inequality gives \eqref{eq:outsens}. On $\Mset$, $e=0$ and $s=0$.
\end{proof}

\bibliographystyle{unsrt}
\bibliography{refs}

\end{document}